\documentclass[aps,prd,showpacs,preprintnumbers]{revtex4}
\usepackage{graphics}
\def \beq{\begin{equation}}
\def \eeq{\end{equation}}
\def \beqa{\begin{eqnarray}}
\def \eeqa{\end{eqnarray}}
\def \tr{{\rm Tr}\,}
\def \det{{\rm Det}\,}

\def \ie{{\sl i.e.\/}}
\def \etal{{\sl et al.\/}}
\def \jhep{{\sl J.\ H.\ E.\ P.\ }}

\def \np{{\sl Nucl.\ Phys.\ }}
\def \pl{{\sl Phys.\ Lett.\ }}
\def \pr{{\sl Phys.\ Rev.\ }}
\def \prl{{\sl Phys.\ Rev.\ Lett.\ }}

\begin{document}
\title{QCD at finite chemical potential\\ with six time slices}
\author{R.\ V.\ \surname{Gavai}}
\email{gavai@tifr.res.in}
\affiliation{Department of Theoretical Physics, Tata Institute of Fundamental
         Research,\\ Homi Bhabha Road, Mumbai 400005, India.}
\author{Sourendu \surname{Gupta}}
\email{sgupta@tifr.res.in}
\affiliation{Department of Theoretical Physics, Tata Institute of Fundamental
         Research,\\ Homi Bhabha Road, Mumbai 400005, India.}

\begin{abstract}
We investigate the Taylor expansion of the baryon number
susceptibility, and hence, pressure, in a series in the baryon chemical potential ($\mu_B$) through
a lattice simulation with light dynamical staggered quarks at a finer
lattice cutoff $a=1/6T$.  We determine the QCD cross over coupling at
$\mu_B=0$. We
find  the radius of convergence of the series at various temeperatures, and
bound the location of the QCD critical point to be
$T^E/T_c\approx0.94$ and $\mu_B^E/T<1.8$. We also investigate
the extrapolation of various susceptibilities and linkages to finite
chemical potential.
\end{abstract}
\pacs{12.38.Aw, 11.15.Ha, 05.70.Fh}
\preprint{TIFR/TH/08-26, hep-lat/yymmnnn}
\maketitle

\section{Introduction}

We report physics obtained in QCD with two light flavours of dynamical
staggered quarks at finite temperature, $T$, and chemical potential,
$\mu$, at lattice spacing $a=1/6T$. We investigate the physics at finite
chemical potential \cite{cep} using the method of Taylor expansions
that was developed in \cite{pressure} and used for QCD earlier in
\cite{nt4,biswa}.  One of the quantities we investigate is the radius
of convergence of the Taylor series, through which we estimate the QCD
critical point. We also investigate the dependence on $\mu$ of various
other quantities of physical interest. Finally, we investigate the linkage
of quantum numbers and its dependence on $T$ and $\mu$. Related earlier
works are \cite{others}.

\begin{figure}
\begin{center}
   \scalebox{0.7}{\includegraphics{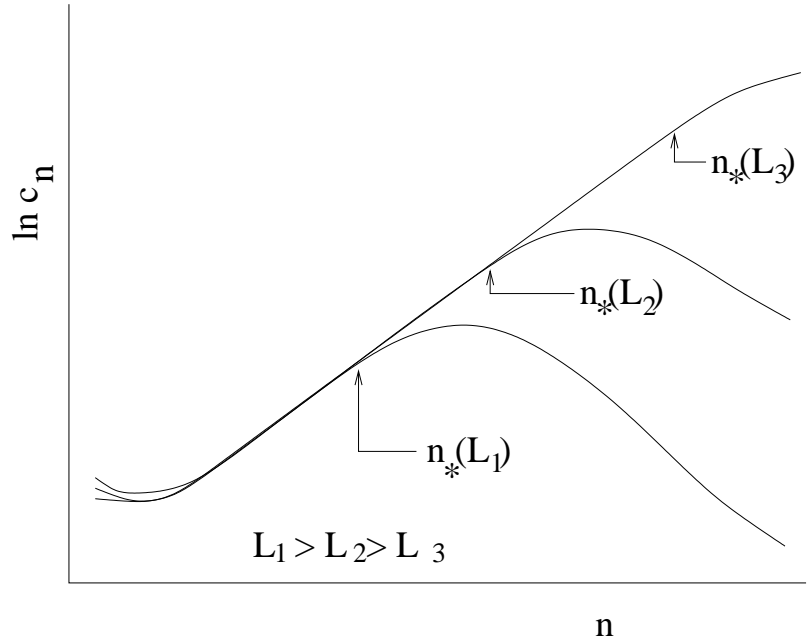}}
\end{center}
\caption{Series coefficients $c_n(L)$ for a quantity that diverges in the
  thermodynamic limit, $L\to\infty$, at the critical point have the finite
  size behaviour shown here. The radius of convergence, $R_n=c_{n-1}/c_n$,
  plateaus for $n<n_*(L)$ before rising to infinity. The case shown here
  corresponds to a singularity on the real axis, since all the $c_n$ in
  the plateau are positive.}
\label{fg.fse}
\end{figure}

That phase transitions are rounded off by finite size effects was
discovered long back by van Hove. The most familiar aspects are seen
when simulations are directly performed in the vicinity of the critical
coupling. Quantities which would diverge in the thermodynamic (infinite
volume) limit are finite.  As a result, a lattice computation never sees
a singularity, but infers its existence from some measures.  Proofs of
the existence usually involve testing extrapolations: such are the main
remaining problems at finite temperature and vanishing chemical potential.
A well-developed finite size scaling theory can be used to study the size,
$L$, dependence of such quantities and extract critical exponents. To
date, the immensity of computational requirements has prevented full
use of this theory for QCD.

The study of the effect of finite size rounding of critical points on
series expansions is, to the contrary, in its infancy. The clearest
fact about such effects is the following: since quantities which should
diverge in the thermodynamic limit merely have finite peaks at finite $L$,
series expansions, strictly speaking, have finite radii of convergence for
finite $L$. In the limit $L\to\infty$ the radius of convergence estimates
the nearest critical point. The mechanism by which this limit is reached
is straightforward. Were one to examine some estimator of the radius of
convergence at order $n$, $R_n(L)$, one would find that up to some $n\le
n_*(L)$ the $R_n(L)$ would approach a finite value.  Such behaviour was
exhibited for the series expansion in QCD in \cite{nt4}.  For larger $n$
the $R_n$ would diverge, in accordance with van Hove's theorem. With
increasing $L$ one would find that $n_*(L)$ approaches infinity. How
the scaling of $n_*(L)$ with $L$ codes for the critical exponent is at
present unknown.

Another question that can be answered by the series expansion is where
the singularity lies. In the case of QCD, the question is whether the
finite radius of convergence of the series expansion in $\mu$ is due to a
singularity at real $\mu$. If it is, then the successive coefficients in
the expansion are positive. At finite $L$ one must examine the signs of
the series coefficients for $n<n_*(L)$. If these are positive, then the
singularity which limits the expansion when $L\to\infty$ is on the real
axis. In \cite{nt4} this argument was used to justify the identification
of the radius of convergence with the critical point. This argument is
also implicit in \cite{misha06}.

In summary, the quark number susceptibilities are Taylor coefficients
in the expansion of the pressure in powers of the chemical potential.
From the series expansion for the pressure, 
\beq
   P(T,\mu_B) = P(T) + \frac12 \chi_B^{(2)}(T) \mu_B^2
      + \frac1{4!} \chi_B^{(4)}(T) \mu_B^4
      + \frac1{6!} \chi_B^{(6)}(T) \mu_B^6
      + \frac1{8!} \chi_B^{(8)}(T) \mu_B^8 + \cdots,
\label{pressure}\eeq
we define the non-linear susceptibilities (NLS) of the $n$-th order,
$\chi_B^{(n)}$. The second order susceptibility, also called the
quark number susceptibility (QNS) \cite{gottlieb}, has the expansion
\beq
   \chi_B(T,\mu_B) = \chi_B^{(2)}(T) + \frac12 \chi_B^{(4)}(T) \mu_B^2
      + \frac1{4!} \chi_B^{(6)}(T) \mu_B^4
      + \frac1{6!} \chi_B^{(8)}(T) \mu_B^6 + \cdots.
\label{chiseries}\eeq
This series is expected to diverge at the QCD critical end point. We define
the radius of convergence of this series as
\beq
   \mu_*^{(n)} = \sqrt{\frac1{n(n-1)}\frac{\chi_B^{(n+2)}}{\chi_B^{(n)}}}.
\label{radcon}\eeq
When successive estimators for $\mu_*^{(n)}$ are equal within statistical
errors to the same value $\mu_*$, we have identified the plateau in the
radius of convergence. This corresponds to the critical point, provided
the singularity in the series occurs at a real value of $\mu_*$. In turn,
this is the case when the coefficients from which the estimates are made
are all positive.

In the next section we present the details of the simulation and the
extraction of the critical coupling. This is followed by a section
in which we report the main results, namely, the extraction of the
QNS up to the eighth order. This results in five terms of the series
for the pressure, and four terms of the series for the baryon number
susceptibility. Using these we report our result for the radius of
convergence of the series, and extract from this our best estimate of the
critical point. In the section after this we discuss the extrapolation
of physical quantities to large chemical potentials.  This extrapolation
throws more light on the nature of the QCD critical point.

Note that the series in eqs.\ (\ref{pressure},\ref{chiseries}) cannot
be continued beyond $\mu_*$ even when all the terms are known
exactly. The truncated series expansion fails even faster. As a result,
it becomes difficult to extrapolate physical quantities to large values
of $\mu_B$. One way to use the series to better advantage is well known:
the method of Pad\'e approximants. The existing theory of Pad\'e
approximants \cite{baker} is adapted to the case where each known
series coefficient has infinite numerical accuracy. When coefficients
are extracted through Monte Carlo estimates, and hence have statistical
errors, new issues arise. We believe that it would be useful to extend
the theory of Pad\'e approximants in this direction.  In the appendix we
make a beginning which is adequate for the purpose of this paper.

\section{Simulations}

\begin{table}
\begin{center}
\begin{tabular}{|l|l|l|rr|rr|rr|}
\hline
$\beta$ & $m_b/T_c$ & $T/T_c$ 
        & \multicolumn{2}{|c}{$6\times12^3$} 
        & \multicolumn{2}{|c}{$6\times18^3$} 
        & \multicolumn{2}{|c|}{$6\times24^3$} \\
 & & & $N_t$ & $\tau_{\rm int}$
   & $N_t$ & $\tau_{\rm int}$
   & $N_t$ & $\tau_{\rm int}$ \\
\hline
5.39  & $0.092\pm0.005$ & $0.89\pm0.01$ & && &&  2284 &  33 \\
5.40  & $0.100\pm0.003$   & $0.92\pm0.01$ 
      & 22599 &  88 & 10099 &  48 &   919 &  35 \\
5.41  & $0.094\pm0.003$ & $0.94\pm0.01$ 
      & 50584 & 197 & && 14580 & 131 \\
5.415 & $0.097\pm0.001$ & $0.97\pm0.01$ 
      & && 17518 & 179 & 14044 & 158 \\
5.42  & $0.099\pm0.001$ & $0.99\pm0.01$ 
      & 39649 & 164 & 35649 & 165 & 27974 & 140 \\
5.425 & $0.1$             & $1.00\pm0.01$
      & 50589 & 189 & 47329 & 214 & 53563 & 267 \\
5.43  && $1.012\pm0.001$   
      & 54619 & 218 & 41349 & 147 & 41869 & 202 \\
5.46  & $0.11\pm0.01$     & $1.21\pm0.01$   
      &   309 &  13 & 10719 &  13 &  1214 &  13 \\
5.54  & $0.10\pm0.01$     & $1.33\pm0.01$   & && &&   969 &   7 \\
5.60  & $0.10\pm0.03$     & $1.48\pm0.03$   & && &&  2891 &   4 \\
5.75  & $0.10\pm0.04$     & $1.92\pm0.05$   & && &&  3626 &   4 \\
\hline
\end{tabular}
\end{center}
\caption{The simulation parameters. These simulations used the R-algorithm
 with a step size of $\delta T_{MD}=0.01$ and a trajectory length of
 $T_{MD}=1$. For tests of accuracy, see later.}
\label{tb.simul}
\end{table}

\begin{table}
\begin{center}
\begin{tabular}{|c|c|c|}
\hline
Operator & $\delta t=0.01$ & $\delta t=0.001$ \\
\hline
$\langle P_s\rangle$ & 1.611 (1) & 1.611 (2) \\
$\langle P_t\rangle$ & 1.611 (1) & 1.611 (2) \\
$\langle {\rm Re\ }L\rangle$ & 0.031 (3) & 0.295 (5) \\
$\langle \overline\psi\psi\rangle$ & 0.293 (9) & 0.291 (9) \\
\hline
\end{tabular}
\end{center}
\caption{Comparison of bulk quantities, namely the spatial and temporal
plaquette averages, the Wilson line and the chiral condensate in runs
with two different MD time steps. In both cases the trajectory length
was 3 MD time units and the first 1002 MD time units were discarded for
thermalization. The number of trajectories used in the comparison was
2916 for the larger time step and 733 for the smaller time step.}
\label{tb.timestep}
\end{table}

\begin{figure}
\begin{center}
   \scalebox{0.6}{\includegraphics{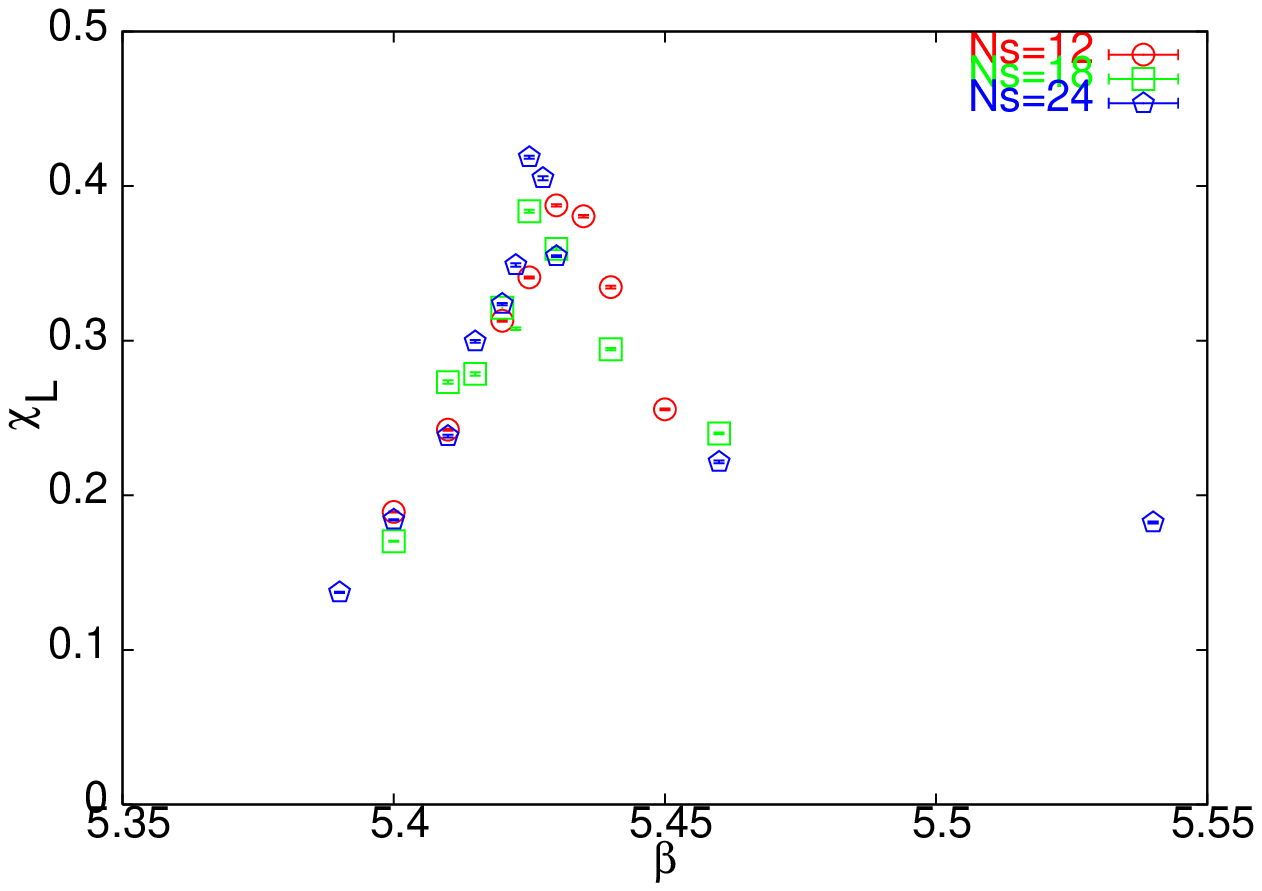}}
   \scalebox{0.6}{\includegraphics{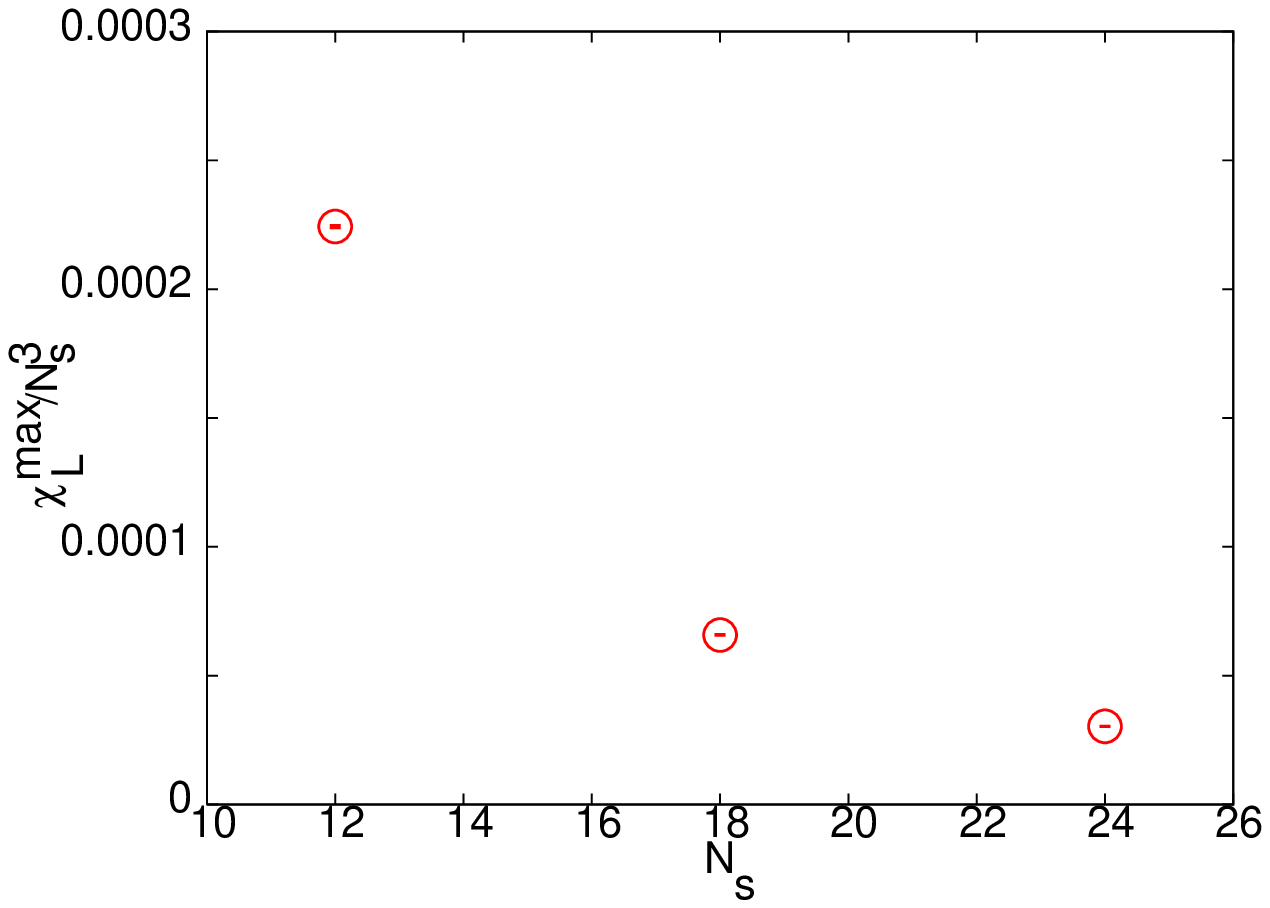}}
\end{center}
\caption{The first figure shows $\chi_L$ as a function of $\beta$. 
 There is some finite size shift in the position of the peak at the lowest
 volumes. The second figure shows the volume dependence of the value of the
 peak, $\chi_L^{max}$ as a function of $N_s$. It is clear that the peak
 grows slower than the volume $N_s^3$.}
\label{fg.chil}
\end{figure}

The simulations were performed using the R-algorithm for hybrid molecular
dynamics. This uses a finite step size, $\delta t$, for the molecular
dynamics. Our main data set is generated using $\delta t=0.01$ and a
total trajectory length of $t=1$ in MD time units. We performed tests
of the accuracy and efficiency of the choices.

Most of our computations were performed with $\delta t=0.01$. This was
found to be adequate for computations at $N_t=4$. We checked our results
at $T/T_c=1.00$ by running a long computation with $\delta t=0.001$
and trajectory length of 3 MD time units. We found complete agreement
between the runs with two different time steps. In Table \ref{tb.timestep}
we show the comparison of bulk quantities computed in the two runs.

Changing the trajectory length from $t=1$ to $t=3$ at
$T/T_c=0.94 \pm0.01$, $1.00\pm0.01$ and $1.92\pm0.05$ did not change
the results for thermodynamic quantities within errors. However,
near $T_c$ the longer trajectories were more effective at reducing
the autocorrelation time. For example, we found that the longest
autocorrelation time at $\beta_c$ was $\tau_{{\rm int}}\approx267$
trajectories for $T=1$, and it reduced to 36 for $t=3$.  As a result, the
CPU time taken to produce decorrelated configurations is reduced by a
factor of about 2.5 on taking the longer trajectories. At $T/T_c=0.94
\pm0.01$ the effective speedup, computed in the same way, was a little
under a factor of two. In the high temperature phase the autocorrelation
times were very small, and there was little to be gained by using longer
molecular dynamics trajectories. There were no changes in thermodynamic
quantities on changing the trajectory length.

A range of $\beta$ was scanned to locate the bare coupling at the finite
temperature crossover, $\beta_c$. The crossover was located by the peak of
the unrenormalized Polyakov loop susceptibility,
\beq
   \chi_L = N_s^3\biggl(\langle L^2\rangle-\langle L\rangle^2\biggr),
    \qquad{\rm where}\qquad L=\frac1{3N_s^3}\sum_x{\rm Re}\tr P(x),
    \qquad{\rm and}\qquad P(x) = \prod_{t=1}^{N_t} U_{\hat t}(x,t).
\label{polsus}\eeq
Here $U_{\hat t}(x,t)$ is the gauge link in the time direction at the
spatial site $x$ on the time slice $t$. We shall show
later that the fourth order quark number susceptibility also peaks at
the same coupling. This is closely related to the inflection point of the
second order susceptibility which is used by various groups
\cite{hotqcd}.

From the peak of $\chi_L$ we identify $\beta_c=5.425\pm0.005$, where the
uncertainty is due to resolution, and not a statistical uncertainty.
There is a little finite volume shift in the position of the peak of
$\chi_L$ at the smallest volume, but no such shift is observed in going
from $N_s=18$ to $N_s=24$ (see Figure \ref{fg.chil}).  While some volume
dependence is visible in the peak of $\chi_L$, with data from just these
three volumes it is not possible to decide whether there is a crossover
or a critical point at $\beta_c$. However, it is possible to rule out
a first order transition, since $\chi_L^{\rm max}/N_s^3$ definitely drops with
increasing $N_s$, as shown in Figure \ref{fg.chil}.

Subsequently, $T=0$ runs were performed on lattices of size $16^4$
and $24^4$ on a grid of $\beta$ to determine the scale. The scale
determination used the value of the plaquette to obtain the renormalized
gauge coupling in the $\overline{\rm MS}$ scheme. The errors in the
scale setting involve the uncertainty in the location of the crossover
coupling, $\beta_c$, statistical errors in plaquette measurements,
and scheme dependence estimated by evaluating the scale also in the E and V
schemes. In the range of temperatures within 20\% of $T_c$, the largest
errors came from the uncertainty in the determination of $\beta_c$.
Better results can only be obtained by using larger spatial volumes. At
larger temperatures, the scheme dependence of the scale set the largest
errors. These can be reduced by going to smaller lattice spacing, \ie,
to larger $N_t$. The scale setting using the crossover for $N_t=6$
is compatible within errors with that obtained earlier for a similar
setting of scales for $N_t=4$.

\section{Quark number susceptibilities}

A quick reminder of our notation \cite{pressure,nt4} is in order. A quark
number susceptibility is obtained by taking a derivative of the pressure
with respect to the chemical potential. In two flavour QCD there are two
possible chemical potentials, $\mu_u$ and $\mu_d$. If one takes $j_u$
derivatives with respect to $\mu_u$ and $j_d$ with $\mu_d$, then the
order of the quark number susceptibility is $n=j_u+j_d$. Since the $u$
and $d$ quarks are degenerate, and indistinguishable at $\mu_u=\mu_d=0$,
we denote the susceptibilities by $\chi_{j_u j_d}$ when $j_u>j_d$ and
$\chi_{j_d j_u}$ when $j_d>j_u$. The susceptibilities are constructed
from expectation values of a string of $\gamma_0$ operators sandwiched
between quark propagators. The operator ${\cal O}_n$ is the operator
with $n$ insertions of $\gamma_0$ into a single fermion loop, and hence
contributes only to $\chi_{n0}$. The operators ${\cal O}_{abc\cdots}$
are products ${\cal O}_a{\cal O}_b\cdots$, and may contribute to several
of the $\chi_{nm}$. The construction of ${\cal O}_n$ on the lattice is
given in detail in \cite{nt4}. We shall discuss results for $\chi_{nm}$
as well as the expectation values $(T/V)\langle{\cal O}_{abc\cdots}\rangle$
(since we discuss only the connected pieces of these expectation values,
we have not used separate notation for that).

The quark number susceptibilities are obtained as expectation values
of fermion loops with various operator insertions \cite{nt4}. These
are evaluated as usual through stochastic estimators. The computations
were optimized using the methods of \cite{nt4}. The need to use
large number of stochastic vectors has been discussed in detail elsewhere
\cite{sewm}. We have taken 500
random vectors for each trace evaluation. With this we are able to control
statistical errors on loops with up to six operator insertions. Even so,
loops with larger numbers of insertions remain noisy.  Thus, at $T_c$
on the largest volume, $\chi_{20}$ gives a signal at 53$\sigma$ and
$\chi_{40}$ at 23$\sigma$, whereas $\chi_{60}$ and $\chi_{80}$ give
signals at $5\sigma$ and $3\sigma$ respectively.  For the two highest
susceptibilities, this level of the signal is an improvement over the
corresponding results with $N_t=4$ at equal $N_s/N_t$.  It was our
experience at coarser lattice spacing that one needs lattices with
larger spatial volumes to control loops with more insertions. We see
this also at the current lattice spacings, at the smallest volumes,
even loops with six insertions are hard to control.

In the following sections we will often compare results for $N_t=4$
and $N_t=6$. These results are meaningful only if they are done holding
other factors fixed. We shall therefore compare the new results obtained
on $6\times24^3$ lattices with those obtained earlier on $4\times16^3$
with the same quark mass, $m/T_c$.

\subsection{Second order}

\begin{figure}
\begin{center}
   \scalebox{0.6}{\includegraphics{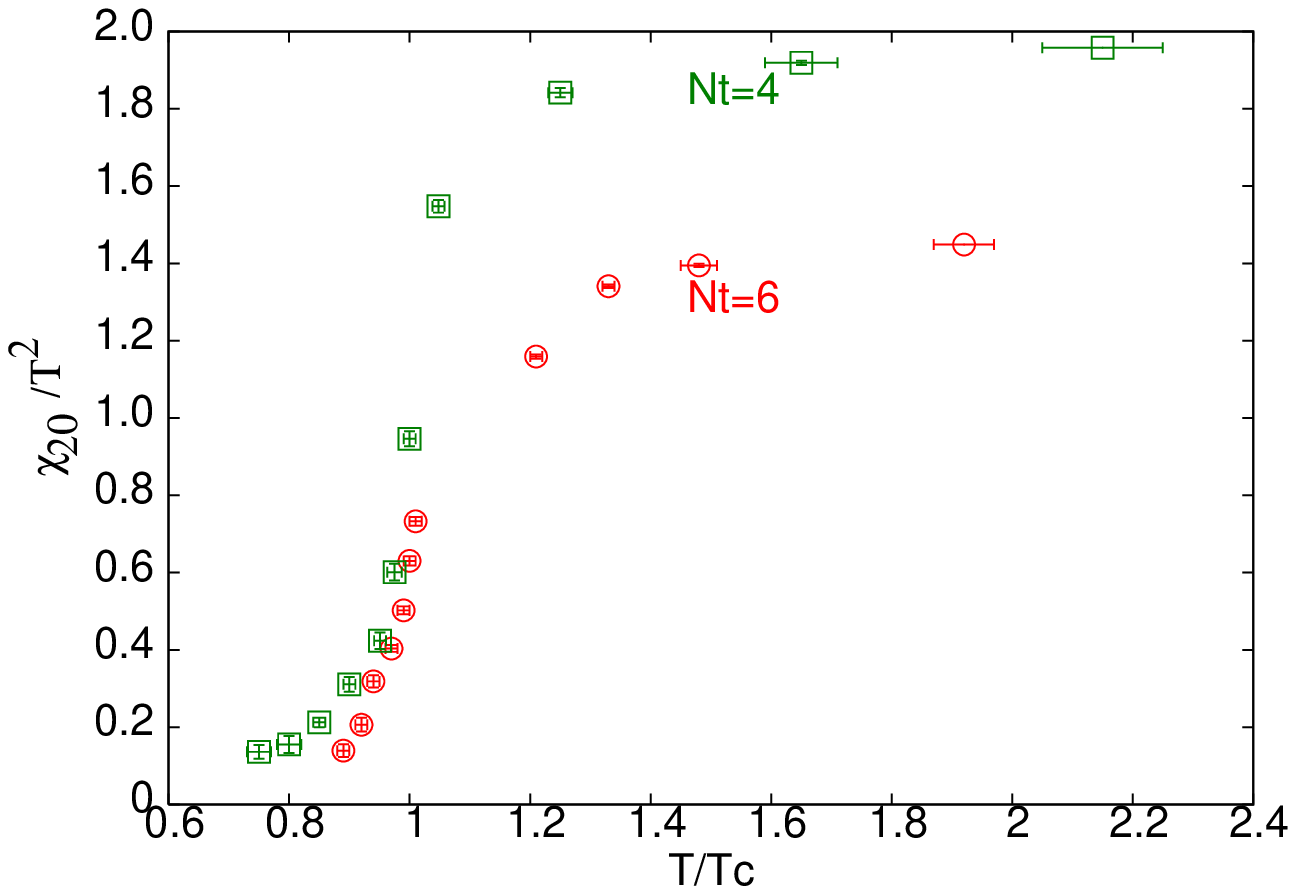}}
   \scalebox{0.6}{\includegraphics{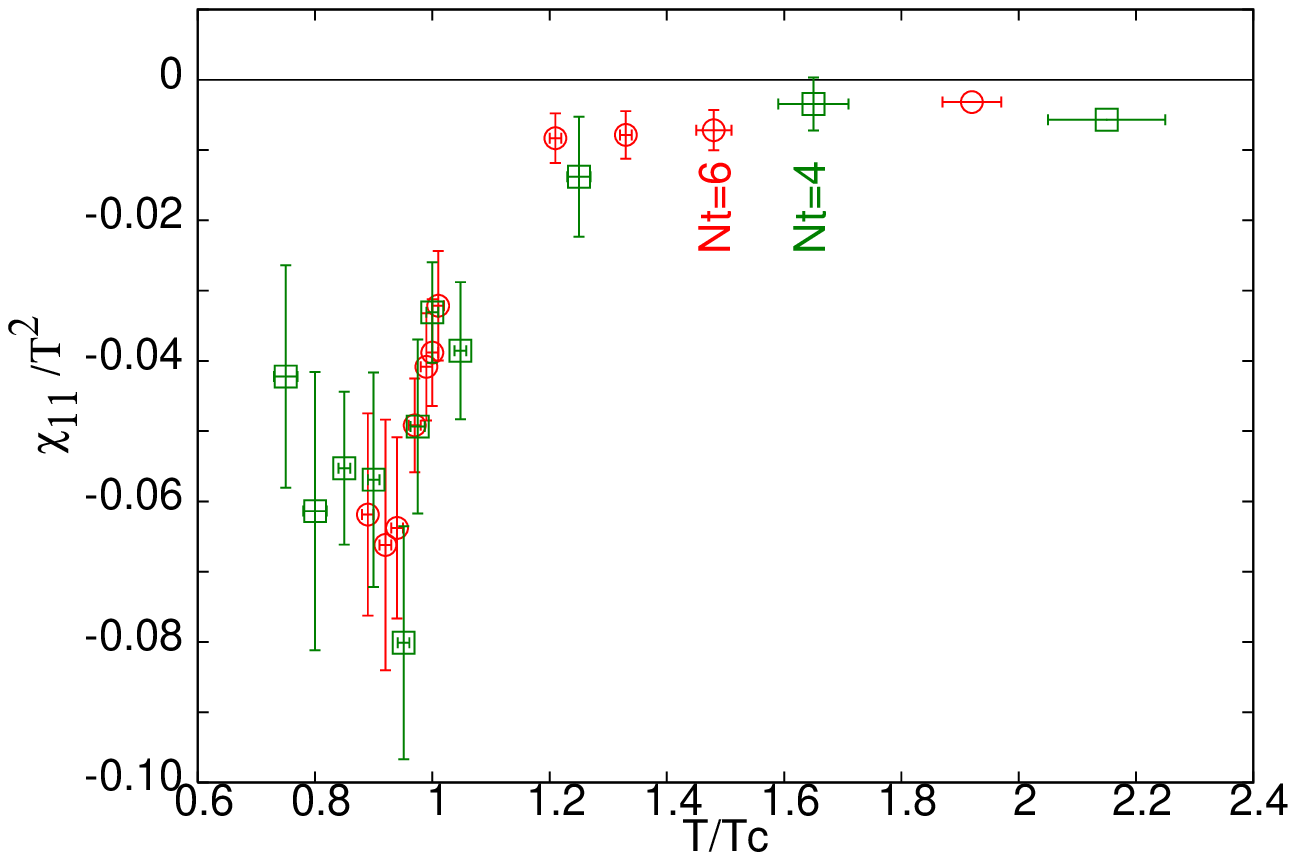}}
\end{center}
\caption{The flavour diagonal and off-diagonal quark number susceptibilities on
 $6\times24^3$ (circles) and $4\times16^4$ (boxes) lattices.}
\label{fg.qns}
\end{figure}

The lowest order quark number susceptibilities are shown in Figure
\ref{fg.qns}. The diagonal susceptibility, $\chi_{20}$ seems to show
significant dependence on $N_t$, \ie, the lattice spacing. This is not
a surprise; after all, even in the quenched theory a similar effect
was seen \cite{quenched}. The off-diagonal susceptibility, $\chi_{11}$
seems to scale better with the lattice spacing.

Note that $\chi_{11}$ takes contributions only from $(T/V) \langle {\cal
O}_{11} \rangle$, whereas $\chi_{20}$ has contributions from this as
well as $(T/V) \langle {\cal O}_2 \rangle$. The results shown in Figure
\ref{fg.qns} indicate that the quark-line disconnected operator has, at
best, marginal lattice spacing dependence. Most of the lattice spacing
dependence seen in $\chi_{20}$ therefore comes from $(T/V) \langle {\cal
O}_2 \rangle$. This last expectation value is the response to a chemical
potential on the isospin component $I_3$ and hence was called $\chi_3$
in some of our early papers. Both this and $\chi_{20}$ change rapidly
near $T_c$ and the ``inflection point'', \ie, the point at which the
slope is maximum can be used as a corroborative measure of $\beta_c$.
Because of the numerical uncertainties in taking derivatives of noisy
data, we will instead use the peak in the fourth order susceptibility.
We discuss these next.

\subsection{Fourth order}

\begin{figure}
\begin{center}
   \scalebox{0.6}{\includegraphics{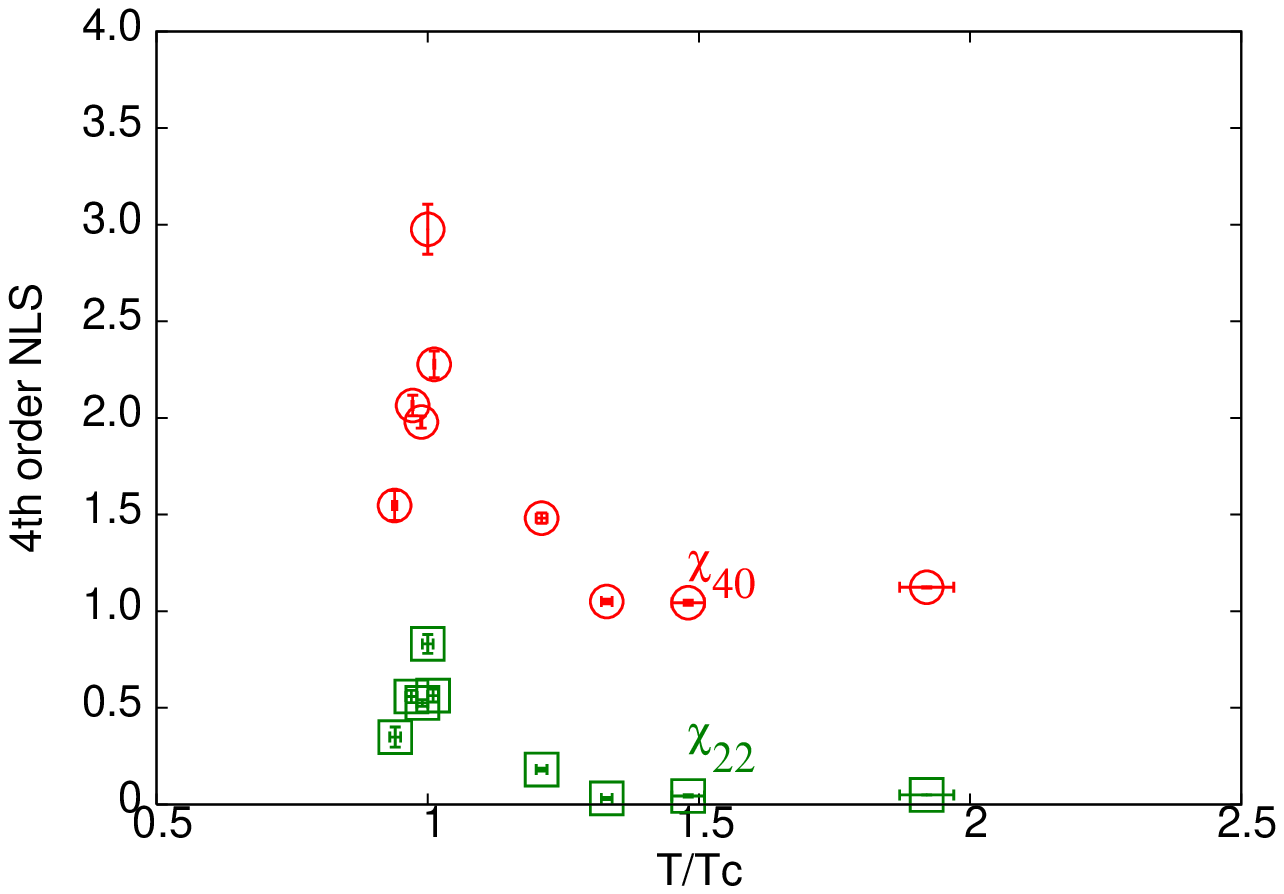}}
   \scalebox{0.6}{\includegraphics{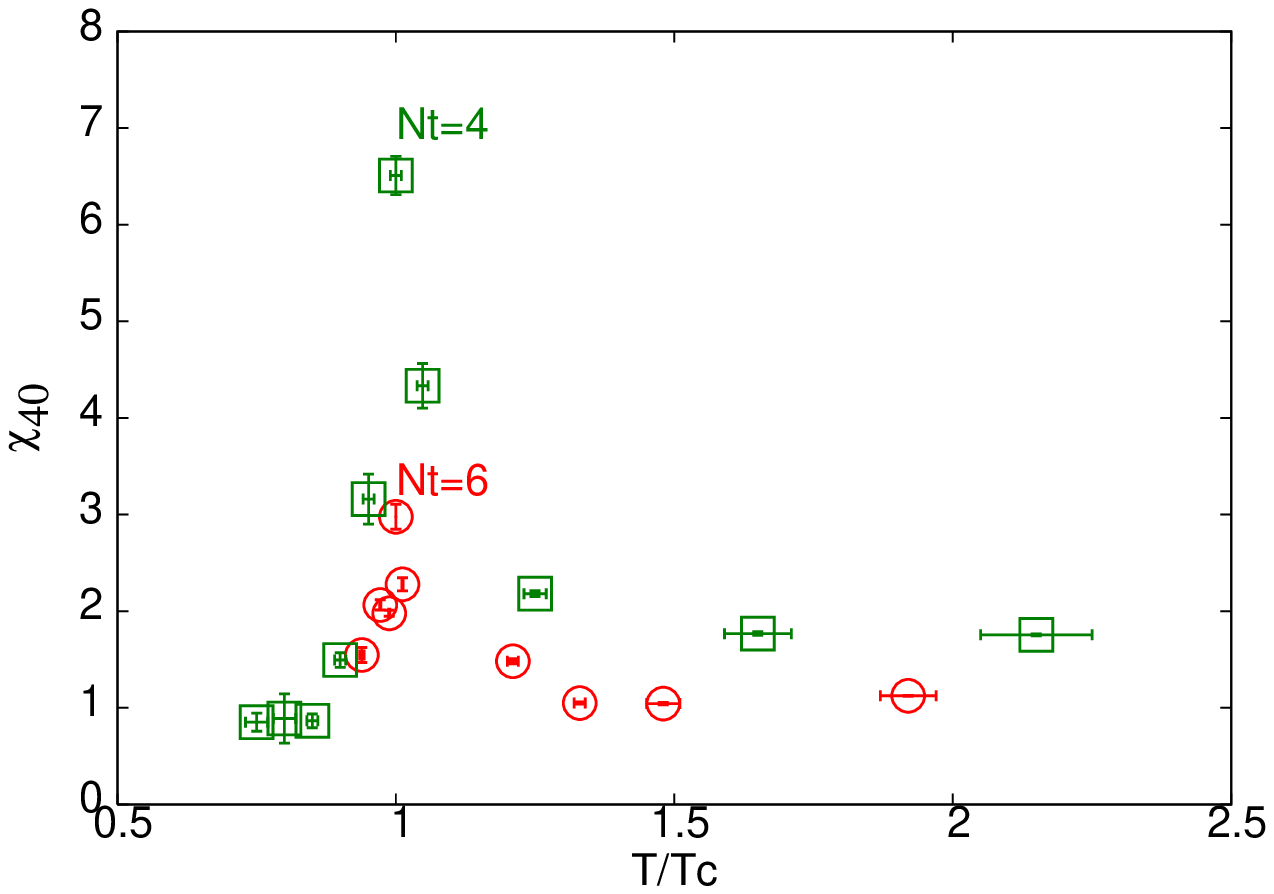}}
\end{center}
\caption{Two of the fourth order susceptibilities on $6\times24^3$ lattices.
  Also shown is a comparison of $\chi_{40}$ on $6\times24^3$ and $4\times16^3$
  lattices.}
\label{fg.nls4}
\end{figure}

\begin{figure}
\begin{center}
   \scalebox{0.6}{\includegraphics{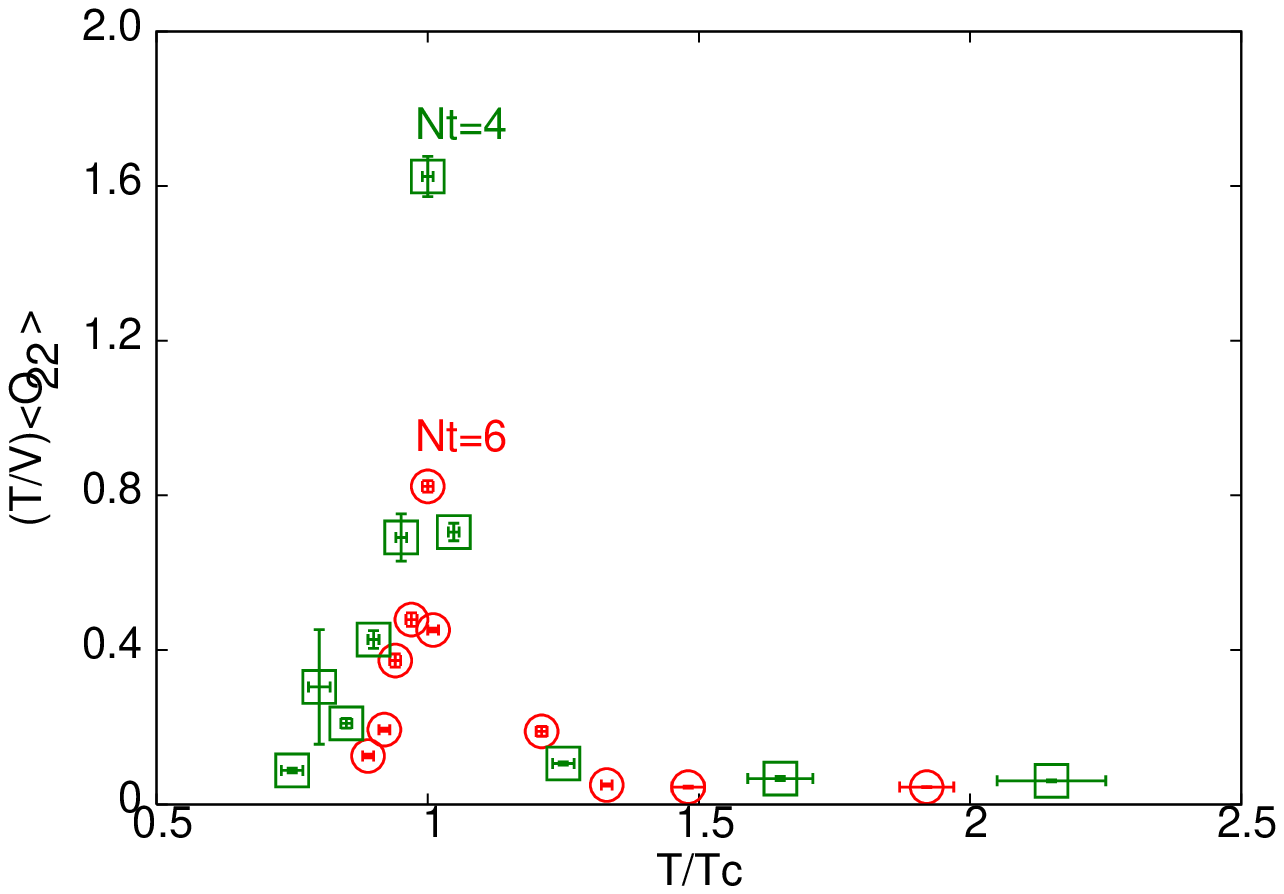}}
   \scalebox{0.6}{\includegraphics{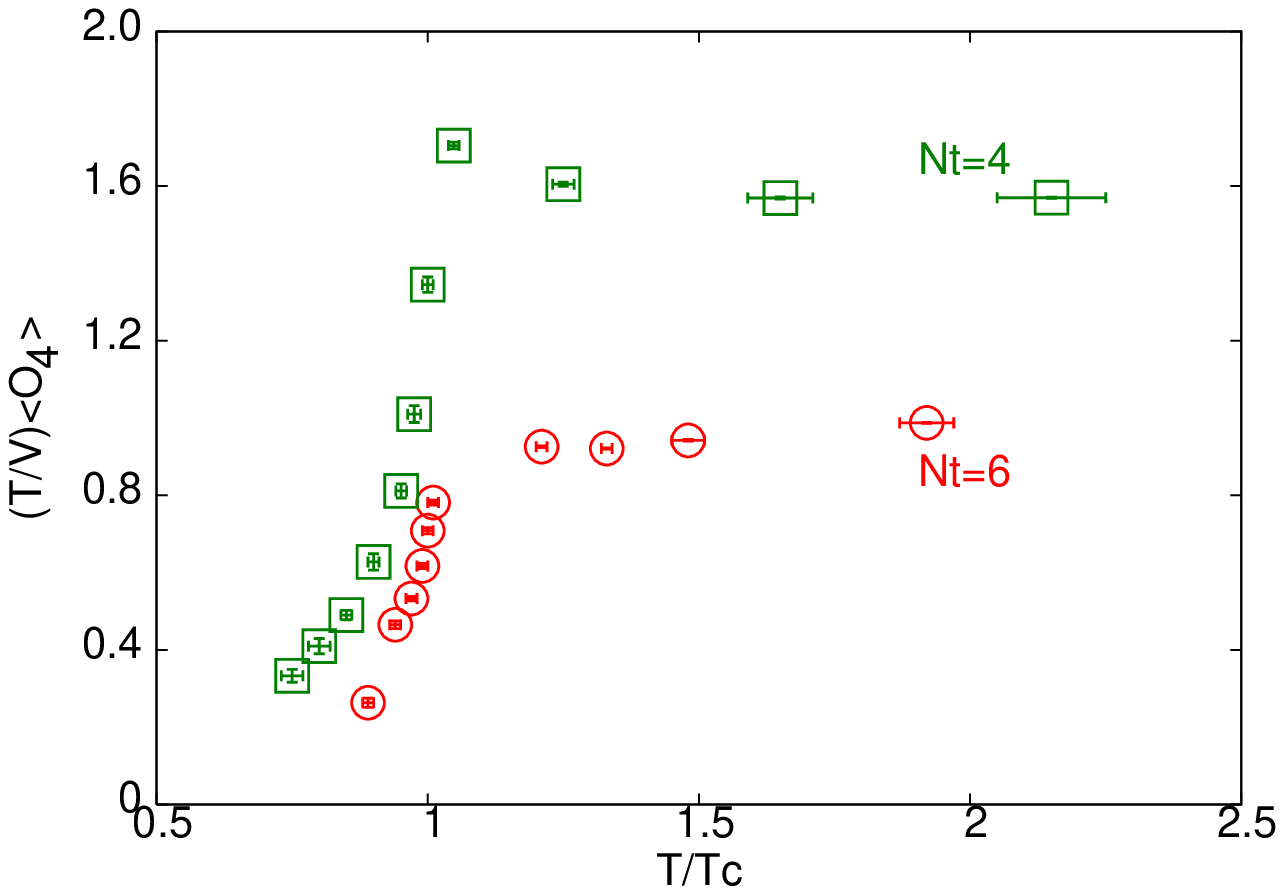}}
\end{center}
\caption{The operator expectation values, $(T/V)\langle {\cal O}_{22}\rangle$ and
  $(T/V)\langle {\cal O}_4\rangle$ on $6\times24^3$ (circles) and $4\times16^3$ 
  (boxes) lattices. The former peaks at $T_c$ whereas the latter exhibits
  a rapid cross over.}
\label{fg.ops4}
\end{figure}

\begin{figure}
\begin{center}
   \scalebox{0.7}{\includegraphics{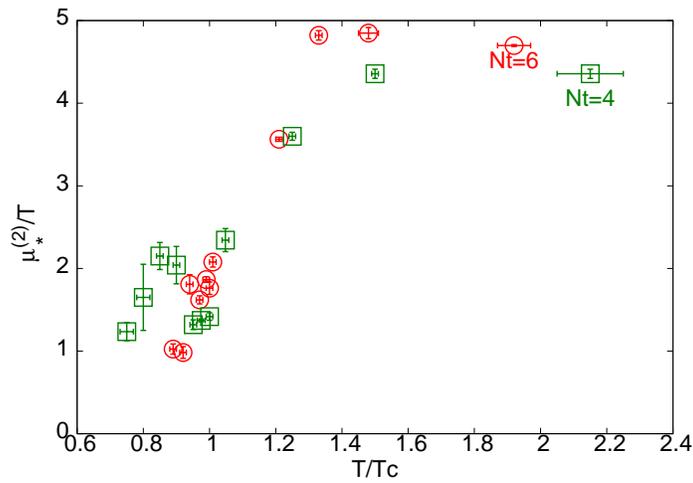}}
\end{center}
\caption{The lowest order estimate for the radius of convergence of the Taylor
  expansion on $6\times24^3$ (circles) and $4\times16^3$ (boxes) lattices.}
\label{fg.rad24}
\end{figure}

Two of the fourth-order susceptibilities are shown in Figure
\ref{fg.nls4}. Both $\chi_{40}$ and $\chi_{22}$ peak at $T_c$. This was
already seen in earlier simulations with $N_t=4$. Within the resolution
of our measurements, we see that the peak in these quantities comes at
exactly the same coupling as the peak in $\chi_L$, at both $N_t=4$ and
6. Like the second order susceptibilities, these too have significant
cutoff dependence. $\chi_{31}$ is much smaller than either of these
susceptibilities and shows no special structure near $T_c$.

The peak at $T_c$ can be resolved into a single operator expectation
value, $(T/V)\langle {\cal O}_{22} \rangle$. This expectation value peaks
at $T_c$ and falls off rapidly on both sides, as shown in Figure
\ref{fg.ops4}. Therefore it can serve as a good measure of the critical
coupling. The expectation value $(T/V)\langle {\cal O}_4 \rangle$, on the other
hand shows a crossover near $T_c$.  One could construct yet another
measure of the critical coupling from the point of steepest slope of
this expectation value, or from its variance, the expectation value
$(T/V)\langle {\cal O}_{44} \rangle$. This last quantity contributes to eighth
order susceptibilities.

Using the fourth and second order quark number susceptibilities, one can
form the first two terms of the series expansion of $\chi_{20}(\mu_B)$
\cite{nt4}. From these coefficients can obtains the lowest order estimate
of the radius of convergence of this series, $\mu_*^{(n)}$. This is shown
as a function of $T/T_c$ in Figure \ref{fg.rad24}. Note that the large
dependence on the lattice spacings seen in each of the susceptibilities
almost cancel out in the estimate of the radius of convergence. The
radius of convergence has smaller dependence on the lattice spacing.

\subsection{Sixth order}

\begin{figure}
\begin{center}
   \scalebox{0.6}{\includegraphics{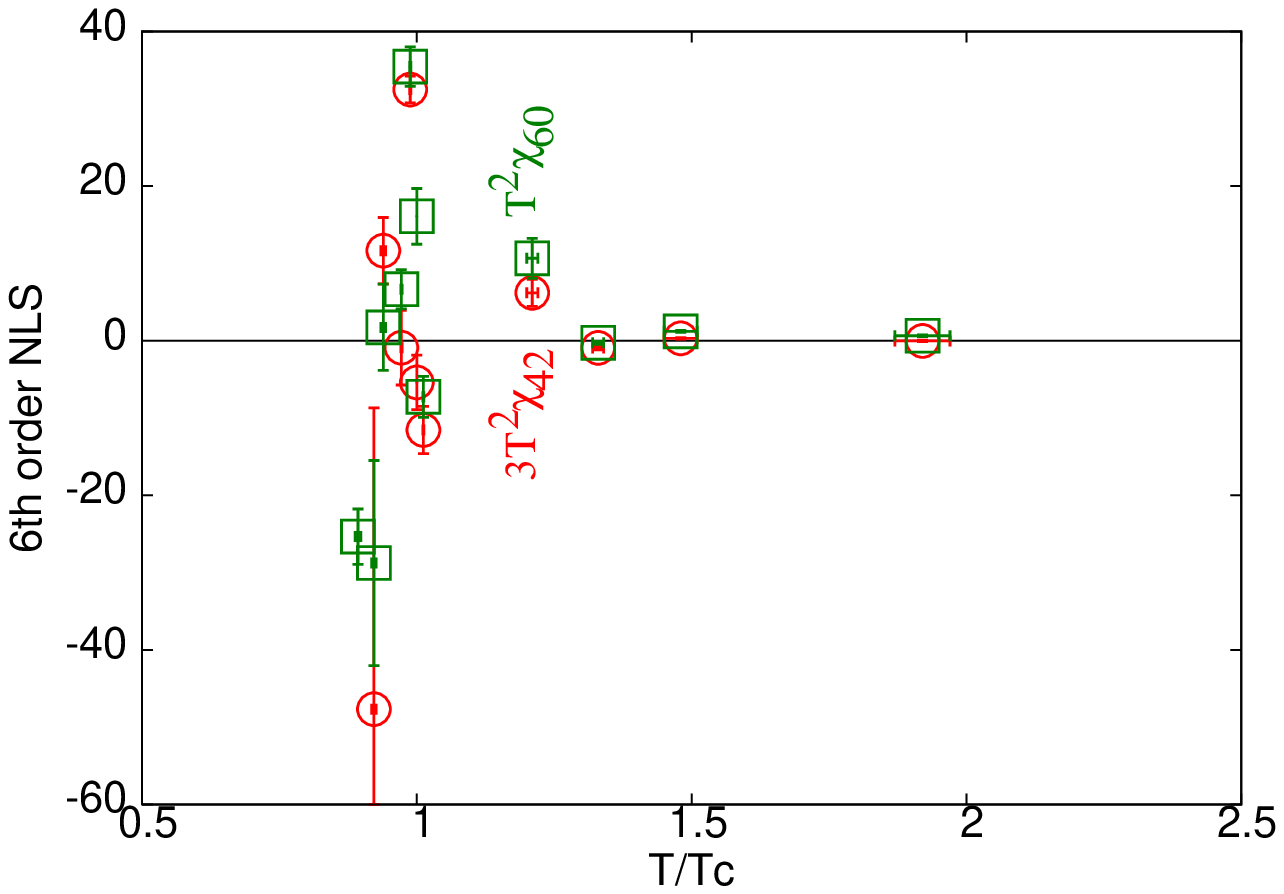}}
   \scalebox{0.6}{\includegraphics{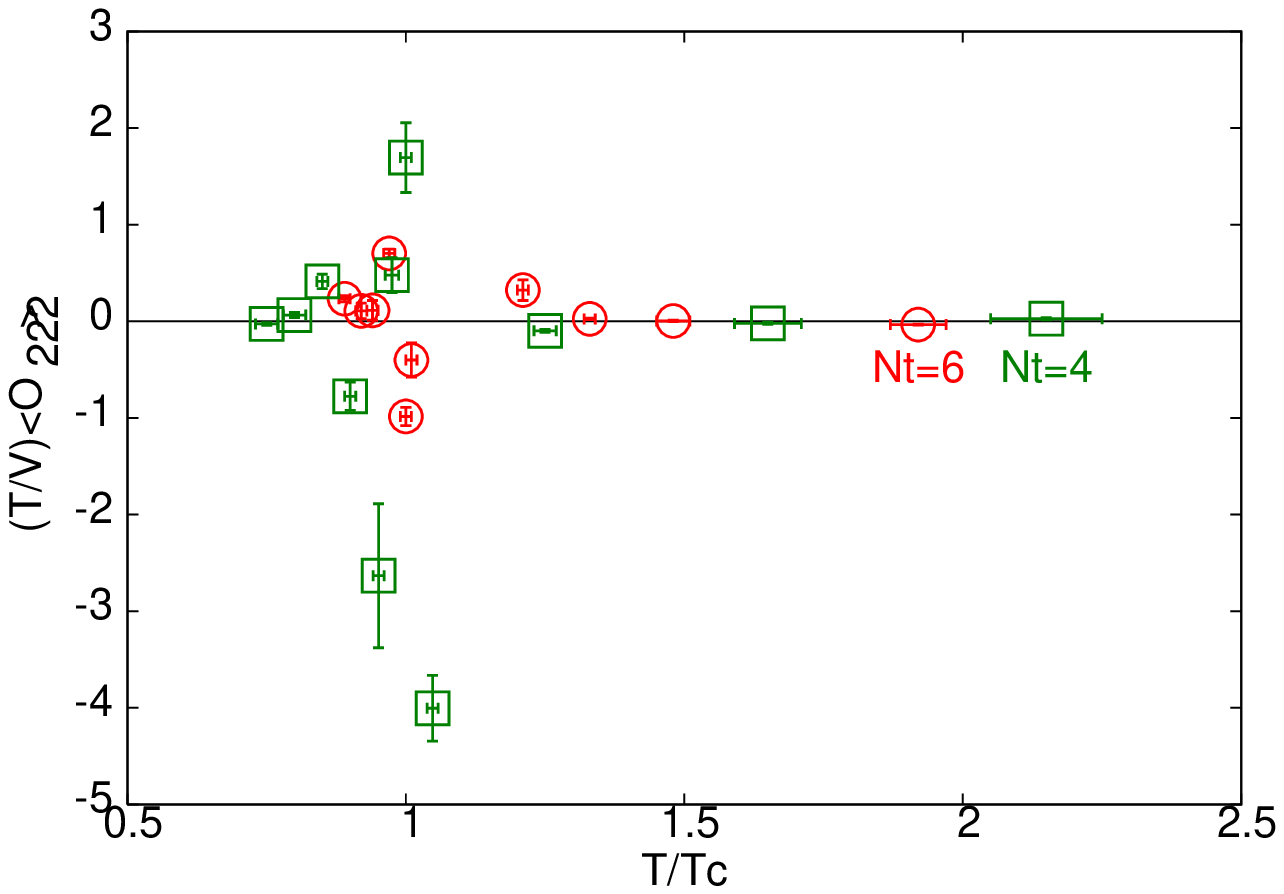}}
\end{center}
\caption{Two of the sixth order susceptibilities on $6\times24^3$ lattices. Also
 shown is the expectation value which determines the shape of both near $T_c$
 on $6\times24^3$ (circles) and $4\times16^3$ (boxes) lattices.}
\label{fg.nls6}
\end{figure}

\begin{figure}
\begin{center}
   \scalebox{0.7}{\includegraphics{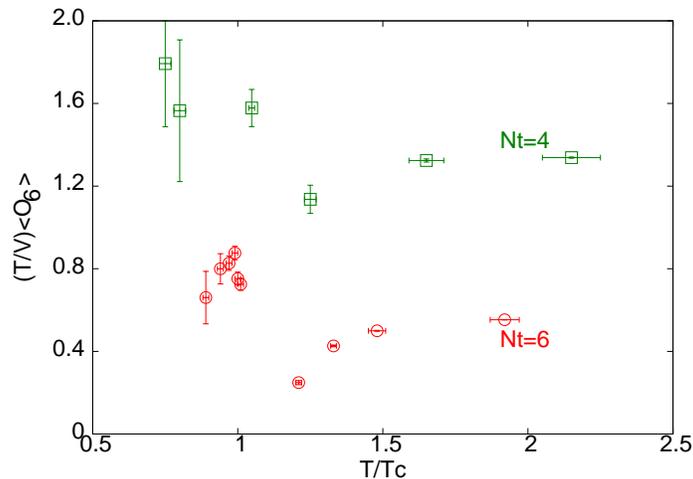}}
\end{center}
\caption{The expectation value of the quark line connected operator which
  contributes to sixth order NLS on $6\times24^3$ (circles) and $4\times16^3$
  (boxes) lattices.}
\label{fg.o6}
\end{figure}

The sixth order NLS is shown in Figure \ref{fg.nls6}. It has been pointed
out earlier that $\chi_{20}$ has the form of a rounded step function, and
that successively higher order NLS have the form of rounded derivatives
of the step function \cite{structure}. For example, the fourth order NLS
has a peak. The sixth order NLS changes sign near $T_c$ and has a maximum
and a minimum flanking the zero. This behaviour is clearly visible in
Figure \ref{fg.nls6}. This peculiar structure comes from the behaviour of
the expectation value $\langle {\cal O}_{222}\rangle$, also shown in Figure
\ref{fg.nls6}. Note that the measurement of $\langle {\cal O}_{222}\rangle$
is noisier than that of $\langle {\cal O}_{22}\rangle$.

The quark-line connected operator expectation value at this order is
$\langle {\cal O}_6\rangle$. This is shown in Figure \ref{fg.o6}. Note
that this has interesting structure below $T_c$ and that the structure
is seen for both $N_t=4$ and 6. As a result, one cannot use $\langle
{\cal O}_6 \rangle$ or its variance for determining $T_c$.

\subsection{Eighth order}

\begin{figure}
\begin{center}
   \scalebox{0.6}{\includegraphics{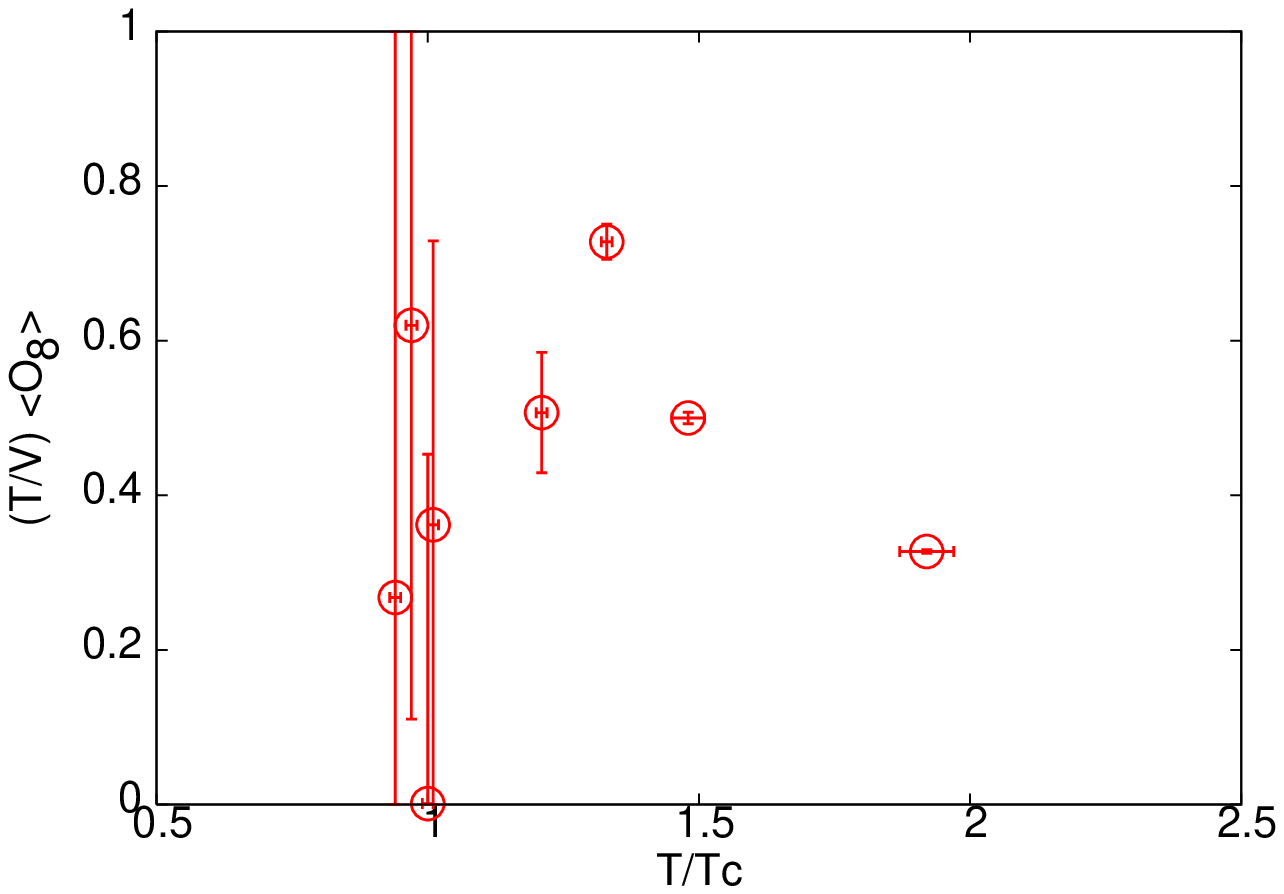}}
   \scalebox{0.6}{\includegraphics{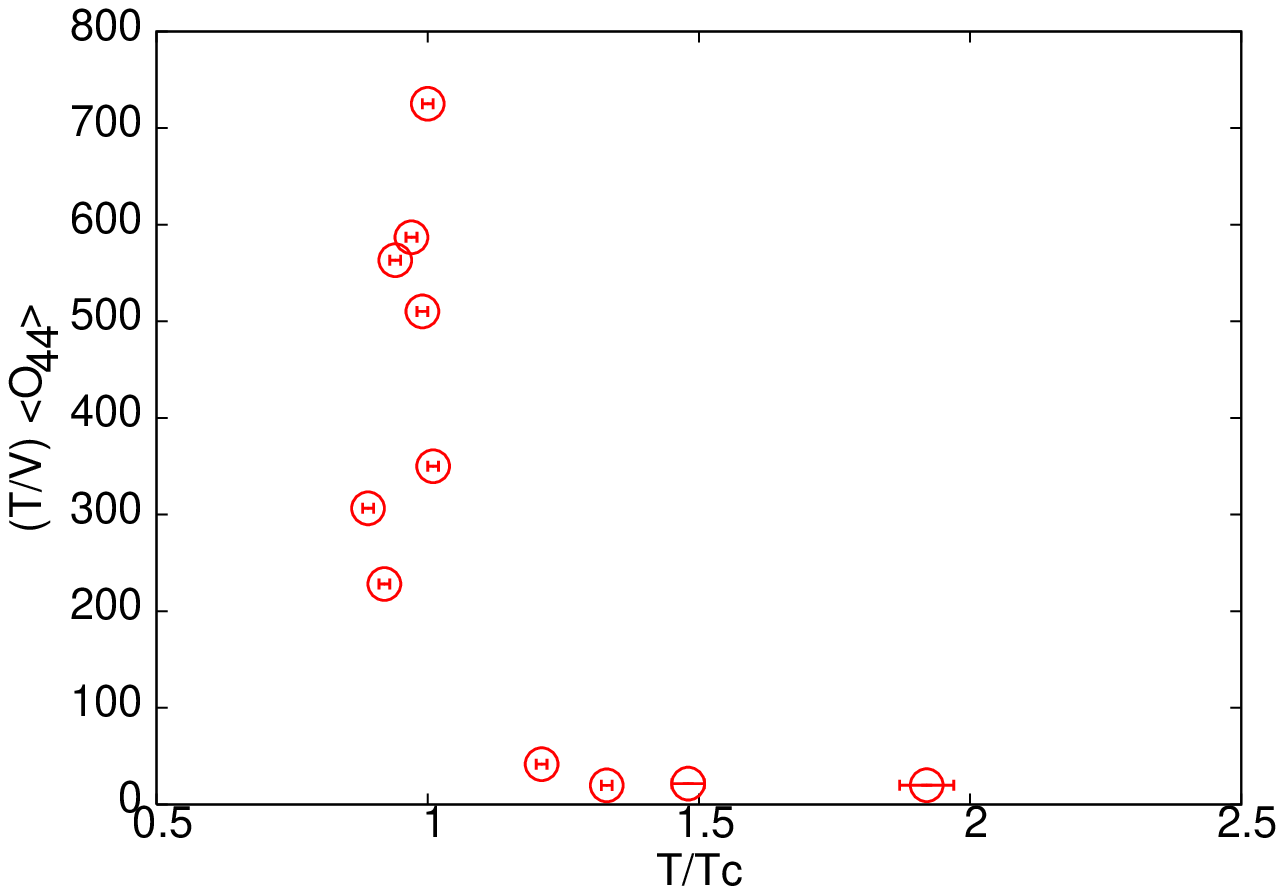}}
\end{center}
\caption{Two of the operators which contribute to the eighth order NLS. The
 quark line connected operator, ${\cal O}_8$ is rather noisy. The connected piece of
 the expectation value of ${\cal O}_{44}$ peaks at $T_c$, as expected, and is the
 least noisy of all operators contributing at this order.}
\label{fg.ops8}
\end{figure}

The eighth order NLS are fairly noisy below and in the vicinity of
$T_c$.  This is partly due to the fact that operators with multiple
quark loops, such as ${\cal O}_{2222}$, become noisier as the number of loops
increases. However, single-loop operators such as ${\cal O}_8$ are also not
under sufficient control at these lattice volumes. We exhibit the
expectation values $\langle {\cal O}_8\rangle$ and $\langle {\cal O}_{44}\rangle$
in Figure \ref{fg.ops8}. Note that there could be structure in the
former below $T_c$, but this is currently obscured by noise. The
latter has a single sharp peak at $T_c$, as argued before, and shows
that $T_c$ identified by the peaks in $\chi_L$, $\chi_{40}$, $\chi_{22}$ and
$(T/V)\langle {\cal O}_{44}\rangle$ are identical within the resolution of
our study.

\subsection{Radius of convergence}

The radius of convergence of the series expansion can be used to estimate
the position of the critical end point of QCD as before. The radius of
convergence gives the distance from the origin where the expansion ceases
to hold. The corresponding singularity lies at real $\mu$ if the coefficients
of the series expansion are all positive. As one comes down in $T$ from large
values of $T$, the series coefficients remain positive down to some lowest
value of $T/T_c=0.94\pm0.01$. At this temperature the radius of convergence
is independent of the order at which it is evaluated and has value
$\mu_B^*/T=1.8\pm0.1$. Thus, our estimate of the position of the critical
end point is
\beq
   \frac{T^E}{T_c} = 0.94\pm0.01,\qquad{\rm and}\qquad
   \frac{\mu_B^E}{T^E} = 1.8\pm0.1.
\label{cep}\eeq
with a lattice spacing of $1/6T$ and a renormalized quark mass that
corresponds to tuning $m_\pi/m_\rho\simeq0.3$, when the spatial size
of the box is $L=4/T$. In comparison, with a lattice spacing of $1/4T$
at the same renormalized quark mass and the same spatial volume, it was
found that $T^E/T_c$ remained unchanged whereas one had $\mu_B^E/T^E =
1.3\pm0.3$. Extrapolation of this result to the thermodynamic limit,
$L\to\infty$ on the coarse lattice yielded an estimate $\mu_B^E/T^E =
1.1\pm0.1$, which, although statistically compatible with the finite
volume result, had a lower mean. It would be interesting to check how
much the new estimate of the QCD critical end point is lowered upon
taking the thermodynamic limit. However, this extrapolation lies outside
the scope of the current work because of the CPU resources needed.

\section{Physics at finite $\mu$}

\subsection{Fluctuations and the quark number susceptibility}

\begin{figure}
\begin{center}
   \scalebox{0.6}{\includegraphics{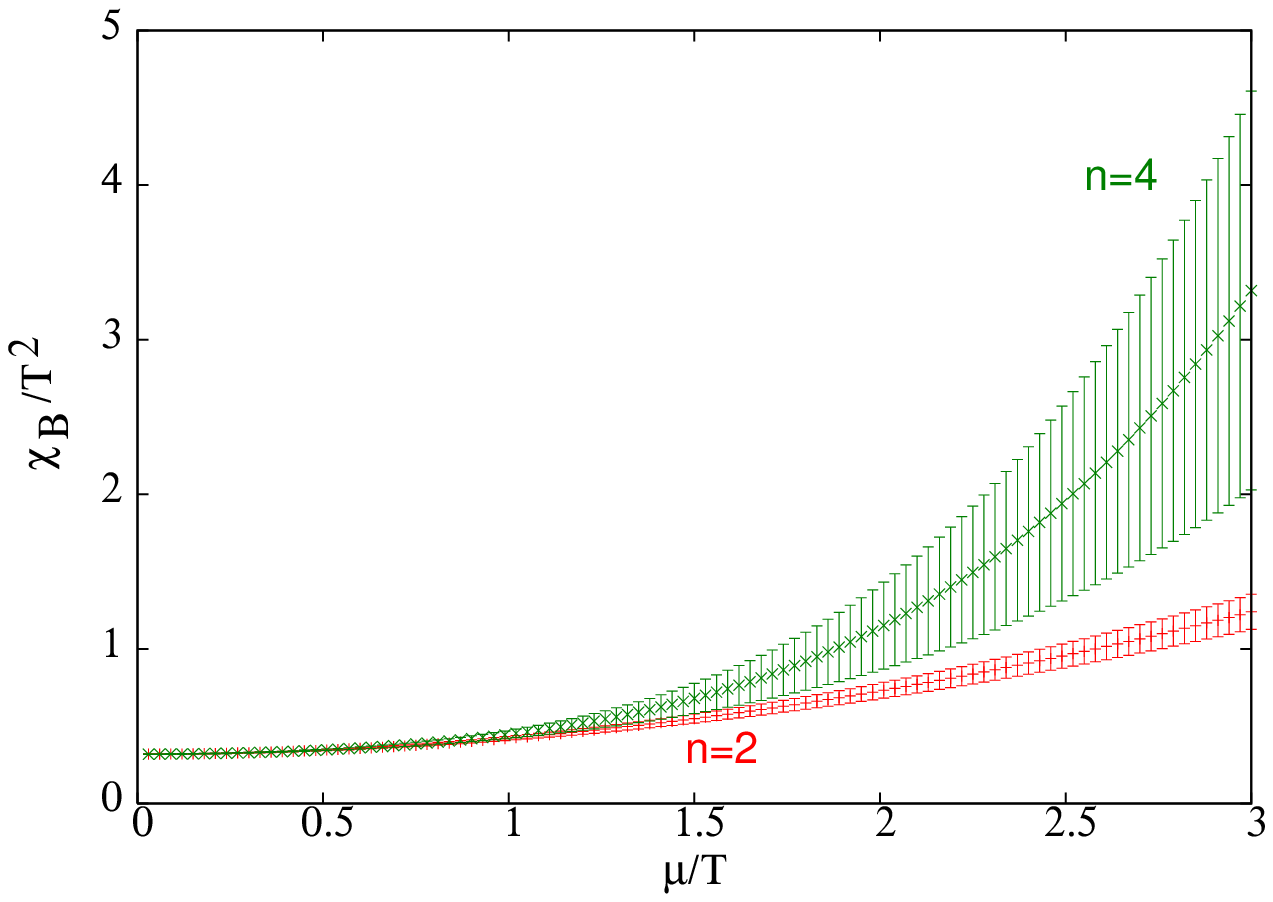}}
   \scalebox{0.6}{\includegraphics{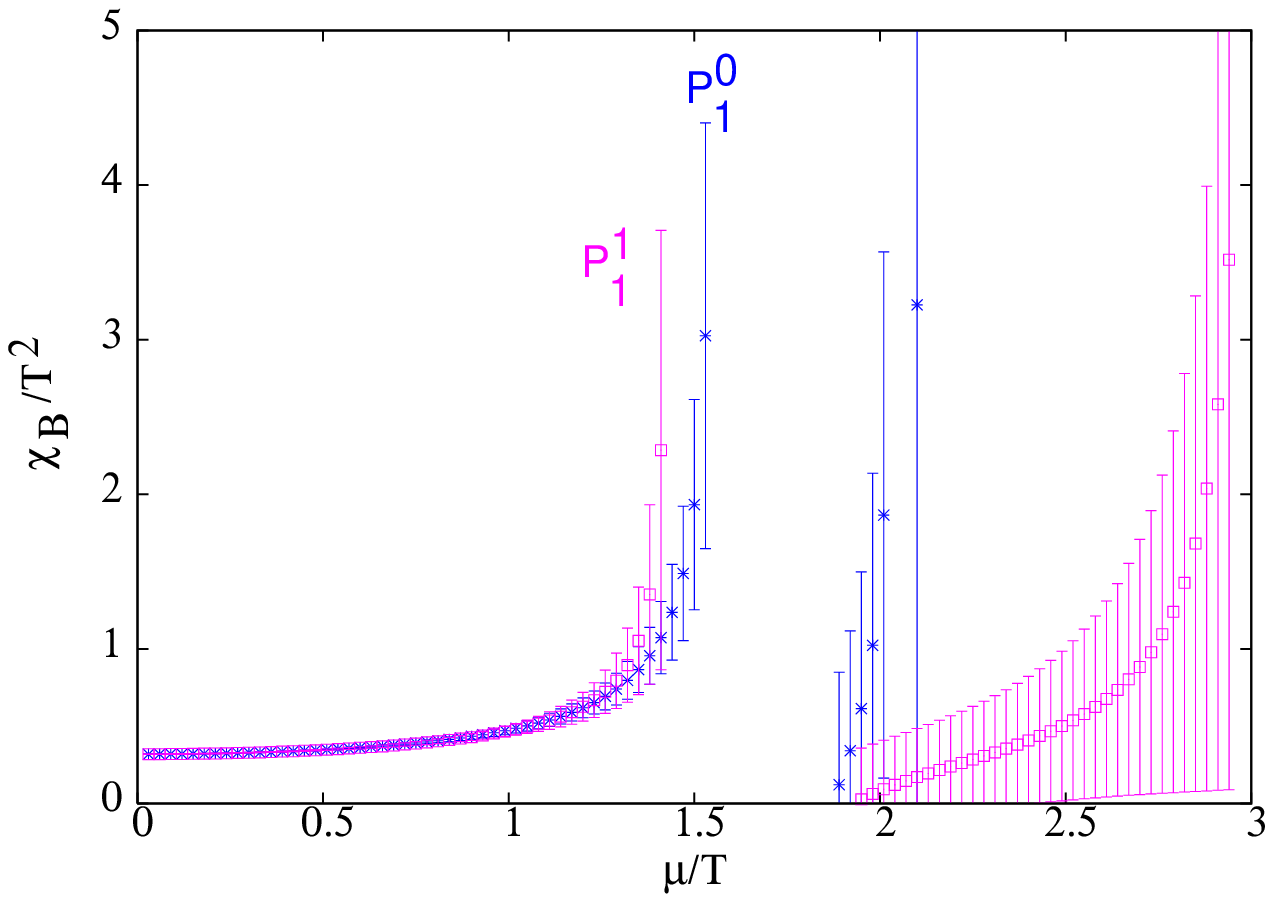}}
\end{center}
\caption{$\chi_B(\mu_B,T^E)/(T^E)^2$ obtained through various extrapolations
  to finite chemical potential. On the left are shown the series expansions to
  orders $n=2$ and $n=4$ (obtained from the 4th and 6th order NLS). The panel
  on the right shows the extrapolations using Pad\'e approximants.}
\label{fg.expansion}
\end{figure}

Baryon number fluctuations, by an amount $\Delta B$ from the expectation
value in a grand canonical ensemble, have a spectrum
\beq
   P(\Delta B) = \exp\left(-\frac{(\Delta B)^2}{2VT\chi_B}\right),
    \qquad{\rm so}\qquad \langle(\Delta B)^2\rangle=VT\chi_B.
\label{fluct}\eeq
It has been proposed that the susceptibilities be measured in
event-to-event fluctuations in heavy-ion collisions \cite{fluct}. Indeed,
the divergence of the width of the spectrum of fluctuations could be
one signal for the detection of the QCD critical point in experiments
\cite{critexpt}. In view of this, it is interesting to estimate $\chi_B$
as a function of $\mu_B$ along the critical isotherm.

While the truncated series expansion can be used to estimate the radius
of convergence, it cannot be used to extrapolate the susceptibility
up to that point. As shown in Figure \ref{fg.expansion}, the series
expansion for $\chi_B(\mu_B,T)/T^2$ taken to orders $n=2$ and $n=4$
fail to agree long before the radius of convergence is reached; nor do
they show any divergence at $\mu_*$. In order to extrapolate the QNS,
one must therefore find more robust techniques.

The usual method is to convert the series to a Pad\'e approximant (see
\cite{mariapade} for a previous application to QCD).  There is extensive
literature on the use of these methods when the series expansion is
known exactly. In Appendix \ref{sc.pade} we extend this theory to
the case relevant to our study, \ie, when the series coefficients are
known only through a Monte Carlo procedure, and hence are known with
certain errors. The appendix examines error propagation in the Pad\'e
approximants, and sets out the basic methods to control these errors. For
our purposes we use the Pad\'e approximants labeled $P^L_1(\mu_B^2/T^2)$
in the notation in Appendix \ref{sc.pade}.

The Pad\'e approximants $P^0_1(\mu^2/T^2)$ and $P^1_1(\mu^2/T^2)$ are shown
in Figure \ref{fg.expansion}. It is interesting to note that they diverge
as $\mu_B^E/T$ is approached. While they disagree with the series expansions
as the radius of convergence is approached, they remain consistent with
each other except very close to the divergence. Note that the errors are
large near the divergence. This seems unavoidable, since any error in the
coefficients will be magnified near the pole. There are also large errors
beyond the pole. It should be possible to control these in future work.

Note that in two flavour QCD one has $\chi_B=2\chi_{BQ}$, at all $\mu_B$,
as long as the isospin chemical potential remains zero. So there are
two independent susceptibilities, $\chi_B$ and $\chi_Q$. In terms of
the previously computed quantities, they are \cite{linkage},
\beq
   \chi_B=\frac29\left(\chi_{20}+\chi_{11}\right),\qquad{\rm and}\qquad
   \chi_Q=\frac2{81}\left(5\chi_{20}-4\chi_{11}\right).
\label{chibchiq}\eeq
It turns out that $\chi_{11}$ remains small within errors even at
larger chemical potential, so that the behaviour exhibited in Figure
\ref{fg.expansion} for $\chi_B$ is also almost quantitatively correct
for $\chi_Q$ after an overall normalization by a factor of $5/9$. In
particular, the divergence in $\chi_B$ is also reflected in $\chi_Q$. This
has consequences which we deal with next.

\subsection{Linkage}\label{sc.linkage}

\begin{figure}
\begin{center}
   \scalebox{0.6}{\includegraphics{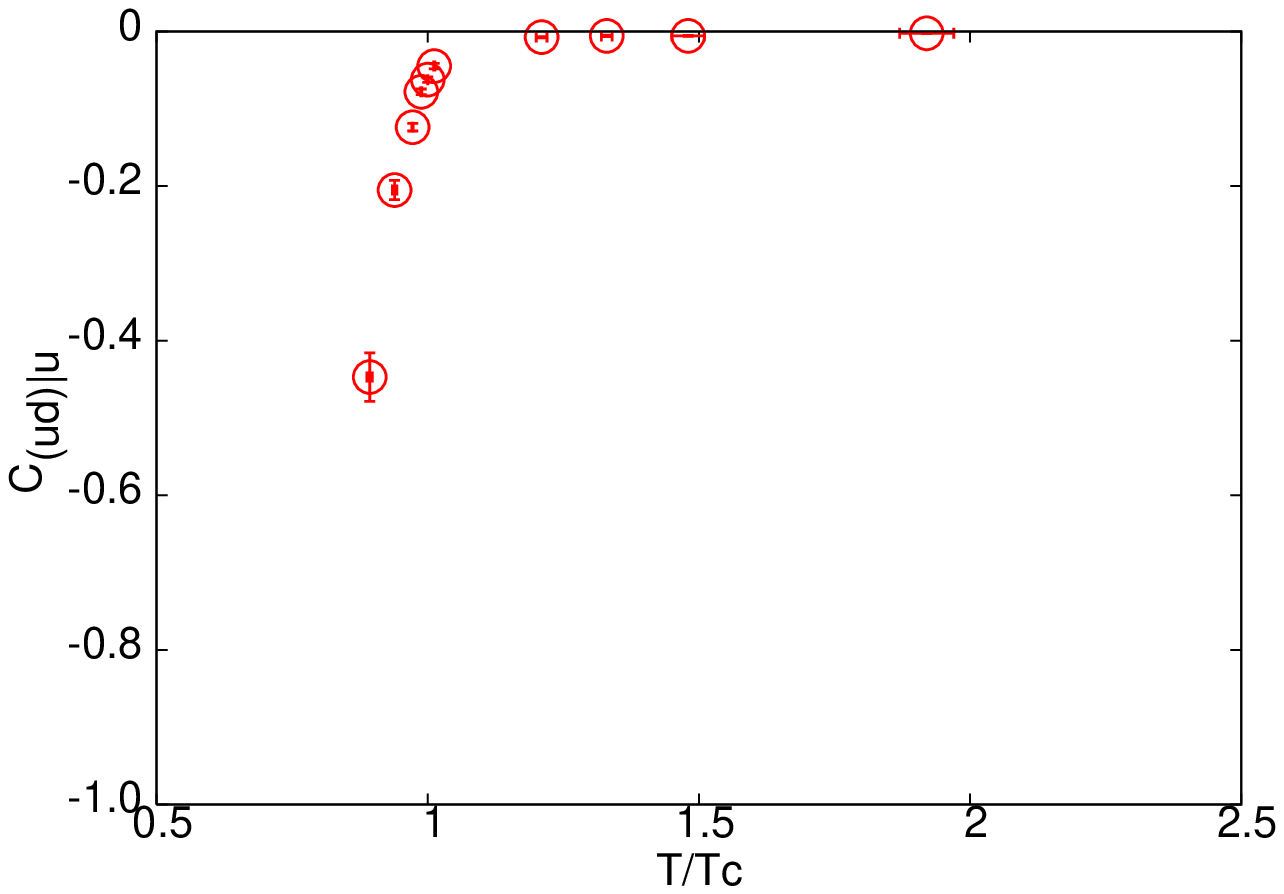}}
   \scalebox{0.6}{\includegraphics{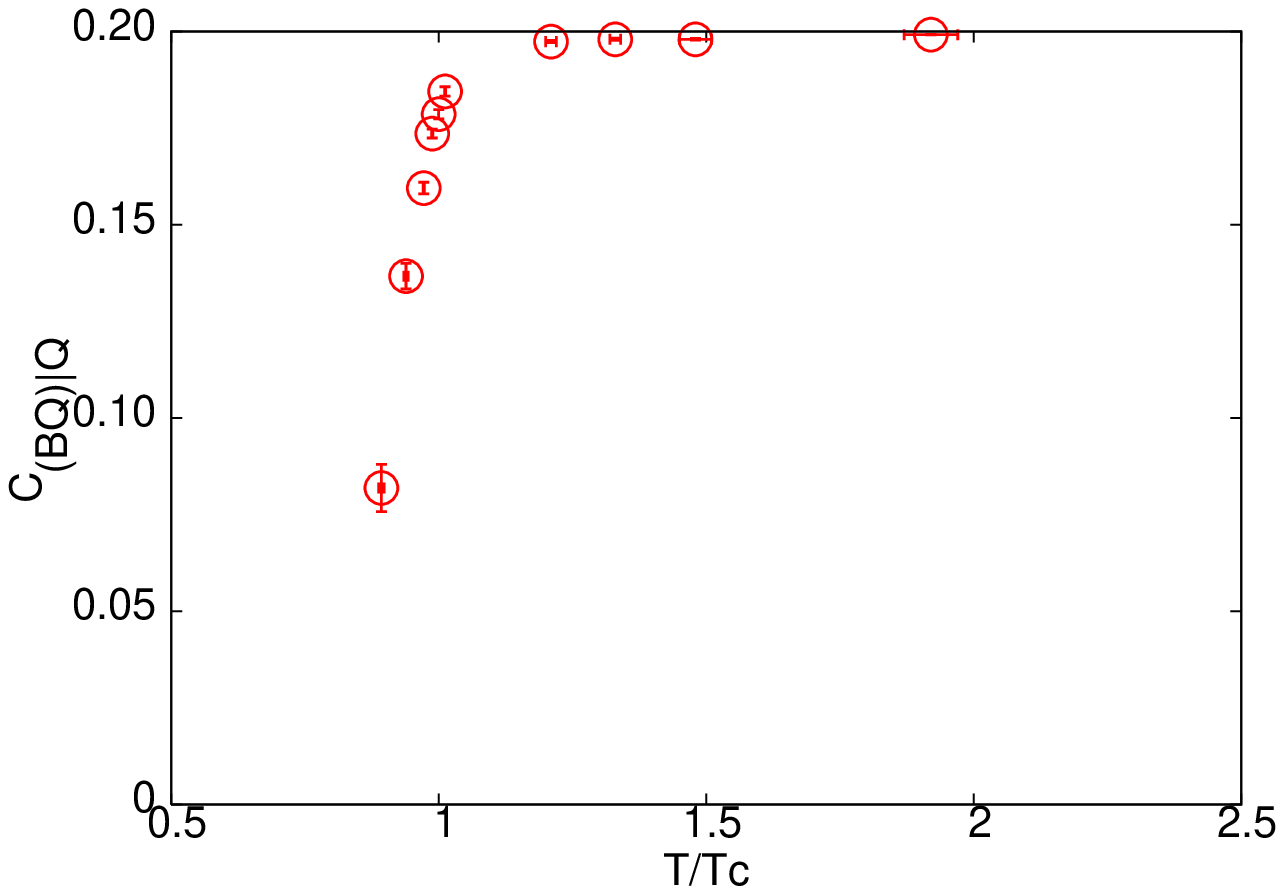}}
\end{center}
\caption{The linkages $C_{(UD)|U}$ (left) and $C_{(BQ)|Q}$ (right) as functions
  of $T/T_c$ at vanishing chemical potentials, as determined on $6\times24^3$
  lattices.}
\label{fg.lnkud}
\end{figure}

\begin{figure}
\begin{center}
   \scalebox{0.6}{\includegraphics{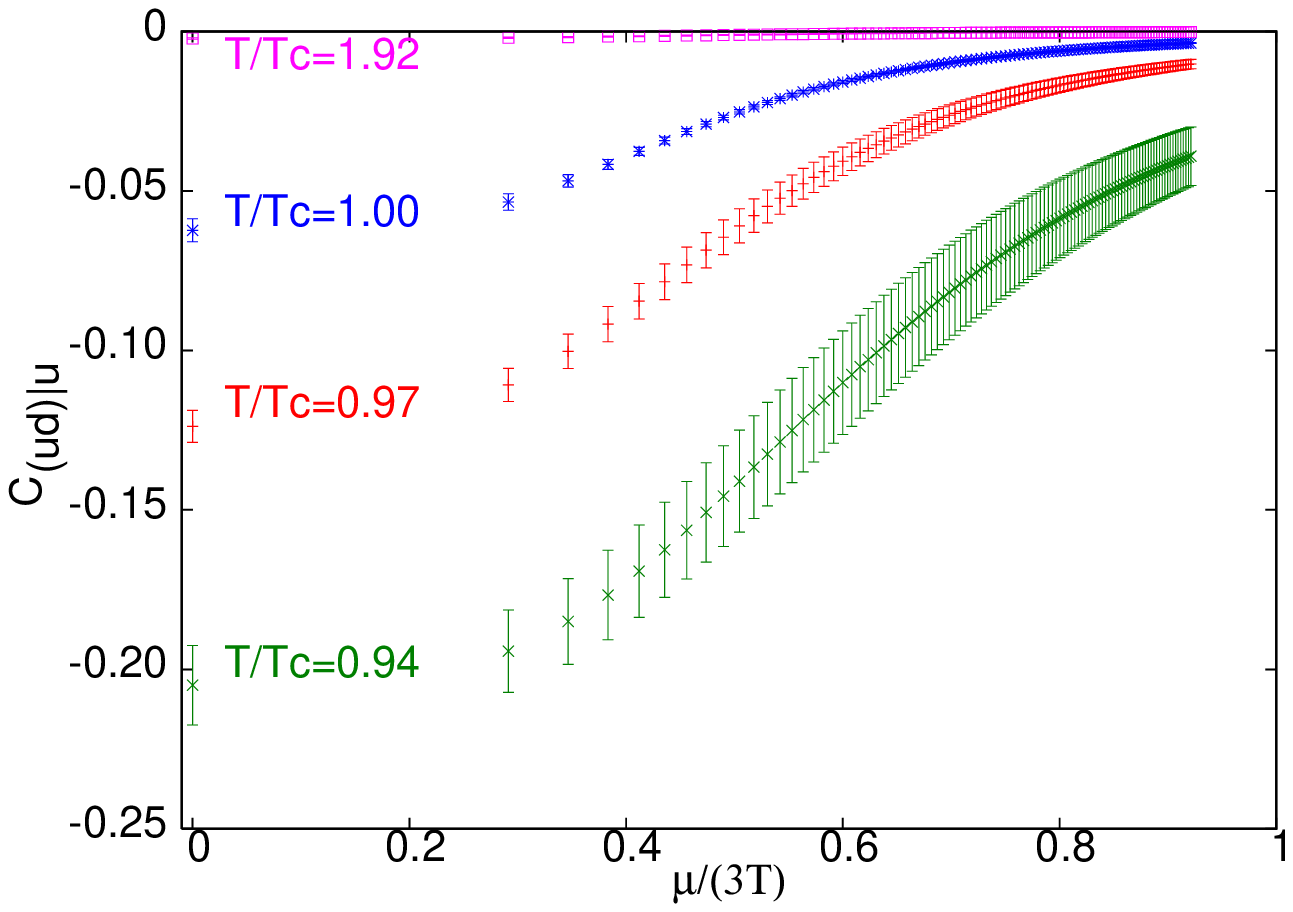}}
   \scalebox{0.6}{\includegraphics{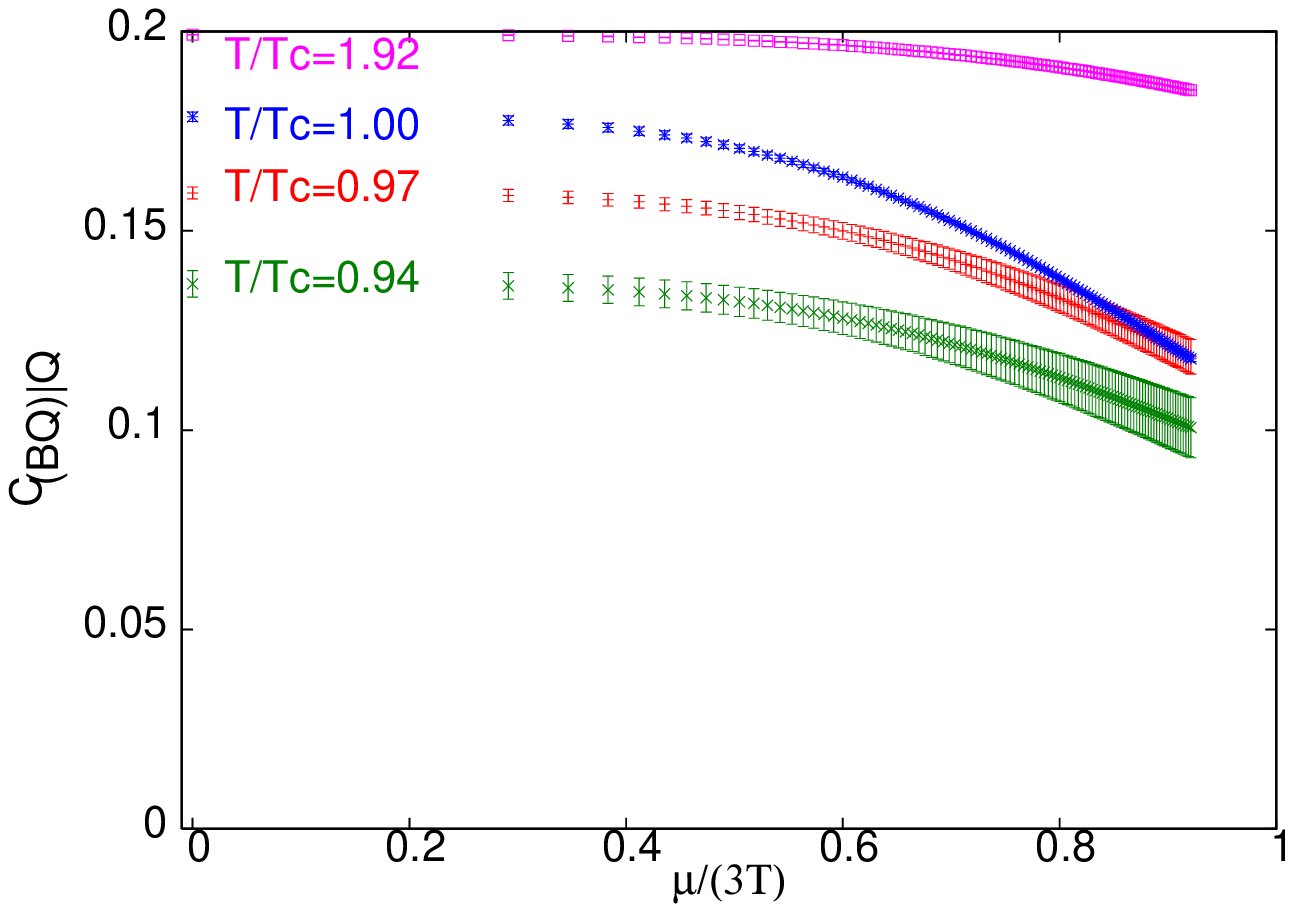}}
\end{center}
\caption{The linkages $C_{(UD)|U}$ (left) and $C_{(BQ)|Q}$ (right) as functions
  of $\mu/T$, as evaluated on $6\times24^3$ lattices, at four different
  temperatures.}
\label{fg.lnkudmu}
\end{figure}

Earlier works have introduced quantities which measure whether two quantum
numbers vary together in thermodynamic fluctuations \cite{koch,linkage}.
The most straightforward measure, called the linkage, utilizes diagonal
and off-diagonal QNS in the form of the ratio
\beq
   C_{(NM)|N} = C_{(MN)|N} = \frac{\chi_{NM}}{\chi_N},
\label{linkage}\eeq
for any two quantum numbers $N$ and $M$. The linkage gives the thermal
averaged amount of the quantum number $M$ excited per unit $N$ in
a thermal fluctuation taking place in the grand canonical ensemble.
In two-flavour QCD one may measure the linkage between U and D quantum
numbers (conventionally +1 for quarks, -1 for antiquarks of the correct
flavour, and zero otherwise). Also related is the linkage between the
baryon number, $B$, and the electrical charge, $Q$.

In Figure \ref{fg.lnkud} we show the temperature dependence of
$C_{(UD)|U}$.  At $T=0$ this quantity should be $-2/3$, since the
lightest excitation is a pion, and the two charged pions each give a
contribution of $-1$, whereas the neutral pion gives a contribution
of 0. In the high-temperature phase, when the lightest excitations are
quark quasi-particles, the linkage should vanish. We see a rapid cross
over between these two regimes, with a very small but non-zero value
being reached at $T_c$. We also exhibit the linkage $C_{(BQ)|Q}$. At
$T=0$ this quantity is expected to vanish, since the lightest charged
particle, the pion, has no baryon number. In an ideal quark gas, this
linkage has value $1/5$. One sees a rapid crossover between these two
values in the vicinity of $T_c$, exactly as for $C_{(UD)|U}$.

In the chiral limit, \ie, when the quark masses vanish, and a second
order phase transition to occur, one would expect that $(T/V) \langle{
\cal O}_{22} \rangle$ becomes infinitely peaked at $T_c$. As a result,
one expects the diagonal susceptibilities to become infinitely sharp,
and the linkages to jump abruptly across the transition. Some part
of the rounding in the linkages is therefore due to the fact that the
quark masses are finite.  However, the rounding of the crossover in the
linkages would be a direct demonstration that there is no abrupt change
from the hadronic to the quark phase: one may use either description over
a small range of temperatures near $T_c$. This could have implications
for the description of hadronization in a fireball, a process which,
at the moment, has a very crude description in terms of the Cooper-Frye
mechanism \cite{cooperfrye}. However, a part of the rounding is also
due to finite volume effects, and it is hard to disentangle the two
in our computation.  It would be an interesting future computation to
understand quantitatively what part of this slow crossover is a finite
volume effect and how much is the effect of a finite quark mass.

Since we have control over the higher order NLS, we can construct a
Taylor series expansion for the linkage and examine its behaviour at
finite chemical potential.  For the analytic continuation of the linkage,
we perform a jack-knife analysis. In each jack-knife bin the Pad\'e
approximant is evaluated at the chemical potential of interest. The
mean of these values is used as the estimator for the continuation,
and the 68\% interval, evaluated non-parametrically, is quoted as the
error bound. The results are shown in Figure \ref{fg.lnkudmu} for several
different temperatures.

At the highest temperature, \ie, $T/T_c\simeq2$, the linkage $C_{(UD)|U}$
is close to zero at $\mu_B=0$ and remains zero for $\mu_B/T\simeq1$. At
temperatures below $T_c$, the linkage is non-zero at $\mu_B=0$ but
evolves smoothly with the chemical potential. For $T>T^E$ we find a
smooth increase with $\mu_B$, the linkage decreasing with larger $\mu_B$.
This illustrates the important point that a finite radius of convergence
for one susceptibility does not imply divergences in other quantities.

Interestingly, the linkage $C_{(BQ)|Q}$ seems to fall marginally with
increasing $\mu_B$. This mild effect can be traced to the fact that
the fourth order coefficient in the Taylor expansion of this linkage is
small and negative. This results in a fall at large $\mu_B$. It would be
useful to check whether this persists at larger volumes, and whether
higher order terms in the expansion turn this around and cause the linkage
to increase.  An interesting alternative possibility is that the fall
in $C_{(BQ)|Q}$ is physical, as is the rise in $C_{(UD)|U}$, and the two
together imply the existence of a phase analogous to the quarkyonic phase
at large $N_c$ \cite{largen}. It is therefore of interest to check these
results further.  Unfortunately, both checks require massive investment
of CPU resources, but are interesting enough that we hope to return to
this soon.

\section{Summary}

We have examined QCD with two flavours of dynamical staggered quarks at
finite temperature with lattice spacing $a=1/6T$ and bare quark masses
tuned to $m/T_c=0.1$. This quark mass is expected to correspond to
$m_\pi/m_\rho=0.3$, and hence our new results are directly comparable to
the older results which were obtained on a coarser lattice with $a=1/4T$
\cite{nt4}. Our simulations were performed on lattices with size $LT\le4$,
where $L$ is the spatial size of the box. We used the R-algorithm with
a step size in MD time units of $0.01$. We have checked that decreasing
this by an order of magnitude to $0.001$ does not change thermodynamic
results (see Table \ref{tb.timestep}). Similarly, we have checked that the
physics results remain unchanged when the trajectory length is changed.

We identified the cross over at vanishing chemical potential through the
Polyakov loop susceptibility, $\chi_L$, (see Figure \ref{fg.chil}) and
then cross checked this through two measures related to the QNS. One
is the peaking of $\chi_{40}$ and $\chi_{22}$
(see Figure \ref{fg.nls4}) which is related to
the ``inflection point'' of the QNS. The other is the peaking of the
operator $(T/V)\langle{ \cal O}_{44} \rangle$, (see Figure \ref{fg.ops8})
which is related to a similar inflection point in $(T/V)\langle{ \cal
O}_4 \rangle$. These measures are consistent with each other within the
accuracy of our computations. The scale setting using this identification
of the cross over is consistent with the earlier scale setting using
coarser lattice spacing \cite{nt4}.

We presented results for the NLS up to eighth order. There are clear
lattice spacing effects, as expected. These are roughly consistent with
earlier determinations of some of these quantities in quenched theory.
While the lattice spacing artifacts for the NLS are very large, sometimes
as much as 100\%, the effect on the radius of convergence is much smaller
(see Figure \ref{fg.rad24}). Our estimate of the critical point of QCD
is based on this radius of convergence. The critical point occurs when
the radius of convergence identifies a singularity on the real axis,
through the fact that the series coefficients are all positive. Caveats
on this are presented in the introduction. Our estimate of the critical
point using finite volume data is (see eq.\ \ref{cep})
\[
   \frac{T^E}{T_c} = 0.94\pm0.01,\qquad{\rm and}\qquad
   \frac{\mu_B^E}{T^E} = 1.8\pm0.1.
\]
This should be compared with our earlier estimate on the same lattice
volume and same renormalized quark mass which gave ${\mu_B^E}/{T^E}
= 1.3\pm0.3$. This is a change of about 26\%, and is statistically
significant. Extension of our results to larger volumes is outside
the scope of this work. In simulations with $a=1/4T$, the estimate of
$\mu_B^E$ dropped by about 16\% on extrapolating to infinite volume.

The series expansion is a good tool for extracting the radius of
convergence, and, through it, the critical point. However, as we show
in Figure \ref{fg.expansion}, it is a bad tool to extrapolate physical
results to high $\mu_B/T$. Pad\'e approximants adjusted to give the
same series expansion seem to perform better, even after taking into
account the propagation of statistical errors. One sees the divergence
of the susceptibility at the critical end point, something that the series
expansion misses altogether.

We also examined the linkages $C_{(UD)|U}$ and $C_{(BQ)|Q}$ (see Figure
\ref{fg.lnkud}). At vanishing chemical potential they show a rapid cross
over from the values expected in the hadronic phase to those expected
for the nearly ideal quarks. The rounding of this transition is closely
related to the question of how sharp the hadronization transition can be
in heavy-ion collisions. A discussion of the issue was presented
in Section \ref{sc.linkage}.

The measurements of the linkages were extended to finite chemical
potential using Pad\'e approximants. Unlike the QNS, they evolve smoothly
through the critical point. Interestingly, on isotherms below $T_c$,
with increasing $\mu_B$, the linkage between U and D quantum numbers
changes towards the ideal quark gas, whereas the linkage between B and
Q changes away from the quark gas. This could indicate the presence of
a quarkyonic phase of QCD matter, although technicalities need to be
sorted out before one can establish this.

The computations were performed over the last two years on the Cray X1
of the Indian Lattice Gauge Theory Initiative (ILGTI) at TIFR. We thank
Ajay Salve for single-handedly taking care of the machine during this
extended period.

\appendix

\section{Pad\'e approximants}\label{sc.pade}

We follow Baker's definition \cite{baker} of a Pad\'e approximant
\cite{pade,gaunt}. The series expansion 
\beq
   f_N(x) = c_0 + c_1 x + \cdots c_N x^N + {\cal O}(x^{N+1})
\label{series}\eeq
evaluated to order $N$ can be used to define the Pad\'e approximant
of order $L/M$,
\beq
   P^L_M[f_N(x)] = \frac{A^L_M(x)}{B^L_M(x)},\qquad
     B^L_M(0)=1,\qquad B^L_M(x) f_N(x)-A^L_M(x)={\cal O}(x^{L+M+1}),
\label{pade}\eeq
where $A^L_M(x)$ and $B^L_M(x)$ are polynomials in $x$ or order up to
$L$ and $M$ respectively. From the matching condition it follows
that $L+M\le N$. Introduce the notation---
\beq
   A^L_M(x)=a_0+a_1 x+\cdots+a_L x^L,\qquad
   B^L_M(x)=1+b_1 x+\cdots+b_M x^M.
\label{notation}\eeq
Then, writing out the matching condition order by order, one obtains
the Pad\'e approximants by solving first for the denominator
\beq
  \left(\matrix{
    c_{L-M+1} & c_{L-M+2} & \cdots & c_L\cr
    c_{L-M+2} & c_{L-M+3} & \cdots & c_{L+1}\cr
    \vdots & \vdots & & \vdots \cr
    c_L & c_{L+1} & \cdots & c_{L+M-1}}\right)
  \left(\matrix{ b_M\cr b_{M-1}\cr \vdots\cr b_1}\right) =
  -\left(\matrix{ c_{L+1}\cr c_{L+2}\cr \vdots\cr c_{L+M}}\right),
\label{denom}\eeq
(with the convention that $c_j=0$ for $j<0$) and then for the numerator
\beqa
\nonumber
   a_0 &=& c_0,\\
\nonumber
   a_1 &=& c_1 + b_1 c_0,\\
   &\vdots& \\
\nonumber
   a_L &=& c_L + \sum_{i=1}^{\min(L,M)}b_i c_{L-i}.
\label{num}\eeqa
The practical importance of Pad\'e approximants arises from the fact
that if the series $f_N$ has a radius of convergence $R$ as $N\to\infty$,
then the series expansion is reliable only for $x<R$, whereas the
Pad\'e approximants can be used for analytic continuation beyond this.
Much of the standard theory
of Pad\'e approximants deals with the cases when the coefficient
matrix in eq.\ (\ref{denom}) has vanishing determinant, and the
information which can then be extracted.

Here we concentrate on a different problem--- that of controlling errors
when the series coefficients are obtained by a Monte Carlo program, and
hence have a given statistical distribution.  We found no discussion of
this in the literature, although it is likely that sporadic attempts
to answer related questions have been made in the past. These
questions become important now that new developments in QCD at finite
chemical potential lead us to analyze series coefficients obtained in
a Monte Carlo process.

When the Pad\'e coefficients are well-defined,
the joint probability distribution
of the series coefficients can be transformed into that of the
coefficients of the Pad\'e approximant using the usual Jacobian
formula---
\beq
   {\cal P}^L_M(a_0,a_1,\cdots,a_L,b_1,b_2,\cdots,b_M) =
   {\cal P}(c_0,c_1,\cdots,c_{L+M}) J,
   \qquad{\rm where}\qquad
   J = \frac{\partial(a_0,a_1,\cdots,a_L,b_1,b_2,\cdots,b_M)}
     {\partial(c_0,c_1,\cdots,c_{L+M})}.
\label{jacob}\eeq

Take the example of $P^0_1$, where $a_0=c_0$ and $b_1=c_1/c_0$,
so that $J=a_0$. Assume that $c_0$ and $c_1$ are drawn from
independent Gaussian distributions of unit mean---
\beq
   {\cal P}^0_1(c_0,c_1) = \frac1{2\pi\sigma_0\sigma_1}
      \exp\left[-\frac12\left\{
         \frac{(c_0-1)^2}{2\sigma_0^2} + \frac{(c_1-1)^2}{2\sigma_1^2}
       \right\}\right].
\label{starting}\eeq
Then the joint distribution of the Pad\'e coefficients can be written down.
The marginal distribution of the Pad\'e coefficient $b_1$, being the
ratio of two Gaussian distributed numbers, is well known \cite{gaussrat}
and given by
\beqa
   {\cal P}^0_1(b_1) &=& {\rm e}^{-\mu(b_1)}\,
      \frac{2+\sqrt{2\pi}\lambda(b_1)}{2\pi\sigma_0\sigma_1\nu^2(b_1)},\\
\nonumber
    {\rm where} && \qquad\nu^2 = \frac1{\sigma_0^2}+\frac{b_1^2}{\sigma_1^2},
  \quad\lambda=\frac1\nu\left(\frac1{\sigma_0^2}+\frac{b_1}{\sigma_1^2}\right),
  \quad\mu=\left(\frac1{\sigma_0^2}+\frac1{\sigma_1^2}\right)-
     \frac{\lambda^2}4.
\label{distb}\eeqa
Note that $\lambda(\pm\infty)=1/\sigma_1^2$, and hence $\mu(\pm\infty)$
is a finite number which depends only on $\sigma_{0,1}$. As a result,
the marginal distribution of $b_1$ is not exponentially damped at
infinity. The power-law damping comes from the factor $1/\nu^2\simeq 1/b_1^2$.
Clearly this distribution has a well-defined expectation value
for $b_1$, but the variance and higher cumulants do not exist.
Thus, statistical measurements of $b_1$ are not subject to the
central limit theorem.

A similar phenomenon occurs with any $P^L_1$. Assume that the series
coefficients are statistically independent and drawn from a Gaussian
of unit mean, then the probability distribution of the Pad\'e coefficients
can be written as
\beq
   {\cal P}^L_1(a_0,a_1,\cdots,a_L,b_1) = 
      \left(\prod_{i=0}^{L+1}\frac1{\sqrt{2\pi\sigma_i}}\right)
    J \exp\left(-\frac{\cal Q}2\right),
\eeq
where $J$ is the Jacobian of the transformation given by $c_i =
\sum_{j\le i} a_{i-j} (-b_1)^j$ for $i\le L$ and $c_{L+1}=c_L b_1$,
and $\cal Q$ is a quadratic form obtained by transforming the arguments
of the Gaussians.

The Jacobian of this transformation is
\beq
   J_L =\left|\matrix{ 1 & 0 & 0 & \cdots & 0 & 0\cr
      -b_1 & 1 & 0 & \cdots & 0 & c_1'(b_1)\cr
      (-b_1)^2 & -b_1 & 1 & \cdots & 0 & c_2'(b_1)\cr
      \vdots & \vdots & \vdots & & \vdots & \vdots\cr
      (-b_1)^L & (-b_1)^{L-1} & (-b_1)^{L-2} &\cdots&1& c_L'(b_1)\cr
     b_1(-b_1)^L & b_1(-b_1)^{L-1} & b_1(-b_1)^{L-2}&\cdots&b_1&c_{L+1}'(b_1)}\right|,
\label{jacobian}\eeq
where $c_i'(b_1)$ is the derivative of $c_i$ with respect to $b_1$. Multiplying
the second row from the bottom by $b_1$ and subtracting that from the last row
gives
\beq
   J_L =\left|\matrix{ 1 & 0 & 0 & \cdots & 0 & 0\cr
      -b_1 & 1 & 0 & \cdots & 0 & c_1'\cr
      (-b_1)^2 & -b_1 & 1 & \cdots & 0 & c_2'\cr
      \vdots & \vdots & \vdots & & \vdots & \vdots\cr
      (-b_1)^L & (-b_1)^{L-1} & (-b_1)^{L-2} &\cdots&1& c_L'\cr
     0 & 0 & 0 &\cdots&0&c_L}\right| = c_L 
        = \sum_{i=1}^L a_{L-i} b_1^i,
\label{jacobianvalue}\eeq
where we have used the relation $c_{L+1}=b_1 c_L$ to write
$c'_{L+1}-b_1 c'_L=c_L$.

The quadratic form in the argument of the exponent can be manipulated
into a particularly useful form by completing the squares---
\beqa
\nonumber
   {\cal Q} &\equiv& \sum_{i=1}^{L+1} \frac{(c_i-1)^2}{\sigma_i^2} =
      {\mathbf a}^TQ{\mathbf a}+2{\mathbf b}^T{\mathbf a}
          + \sum_{i=1}^{L+1}\frac1{\sigma_i^2}
    = ({\mathbf a-\overline a})^TQ({\mathbf a-\overline a}) +\mu,\\
   {\rm where} && \mu = \sum_{i=1}^{L+1}\frac1{\sigma_i^2}
       -\overline{\mathbf a}^T Q\overline{\mathbf a} \quad{\rm and}\quad
    \overline{\mathbf a} = Q^{-1} {\mathbf b},
\label{quadratic}\eeqa
where the real symmetric matrix $Q$ and the vector ${\mathbf b}$
can be easily written down. We do this next for the special case when
all the $\sigma_i$ are equal to $\sigma$.

Define the sequence of polynomials
\beqa
\nonumber
   p_j(b_1) &=& 1+b_1^2+b_1^4+\cdots+b_1^{2j} = 1+b_1^2 p_{j-1}(b_1),\\
   q_j(b_1) &=& 1-b_1+b_1^2-\cdots+(-b_1)^j = 1-b_1 q_{j-1}(b_1).
\eeqa
In terms of these, one writes
\beq
   Q = \frac1{\sigma^2}\left(\matrix{ 
    p_{L+1} & -b_1 p_L & b_1^2 p_{L-1} & \cdots\cr
    -b_1 p_L & p_L & -b_1 p_{L-1} & \cdots\cr
    b_1^2 p_{L-1} & -b_1 p_{L-1} & p_{L-1} & \cdots\cr
    \vdots & \vdots & \vdots & }\right),\qquad
   {\mathbf b} = \left(\matrix{q_L \cr q_{L-1} \cr q_{L-2} \cr \cdots}\right).
\eeq
In order to find the determinant of $Q$, we do the following row
operations--- starting from the top, add to each row $b_1$ times
the next row. This reduces the determinant to a lower triangular form
\beq
   \det Q = \frac1{\sigma^2}\left|\matrix{
     1 & 0 & 0 & \cdots & 0\cr
     -b_1 & 1 & 0 & \cdots & 0\cr
     b_1^2 & -b_1 & 1 & \cdots & 0\cr
     \vdots & \vdots & \vdots & & \vdots \cr
     (-b_1)^L p_1 & (-b_1)^{L-1} p_1 & (-b_1)^{L-2} p_1 & \cdots & p_1}
    \right| = \frac{p_1(b_1)}{\sigma^2}.
\label{detq}\eeq

The solution of the equation $Q\overline{\mathbf a}={\mathbf b}$
can be obtained by the same operations.  They yield the reduced
equation
\beq
  \left(\matrix{
     1 & 0 & 0 & \cdots & 0\cr
     -b_1 & 1 & 0 & \cdots & 0\cr
     b_1^2 & -b_1 & 1 & \cdots & 0\cr
     \vdots & \vdots & \vdots & & \vdots \cr
     (-b_1)^L p_1 & (-b_1)^{L-1} p_1 & (-b_1)^{L-2} p_1 & \cdots & p_1}
    \right)\overline{\mathbf a} = \sigma^2
    \left(\matrix{1\cr 1\cr 1\cr \vdots\cr 1}\right),
\eeq
where we have used the relation $q_i(x)=1-x q_{i-1}(x)$, to reduce the
vector $\mathbf b$. This gives---
\beq
   \overline{\mathbf a} = \sigma^2 \left(\matrix{1\cr
     1+b_1\cr 1+b_1\cr \vdots\cr 1+b_1\cr \frac1{1+b_1^2}+b_1}\right).
\eeq
Finally,
\beq
   \mu=\frac{L+1}{\sigma^2} - {\mathbf b}^T\overline{\mathbf a}
     = \frac{L+1}{\sigma^2} - \sum_{i=1}^L[q_i(b_1)+b_1 q_{i-1}(b_1)]
       -\frac1{1+b_1^2}.
\eeq
Clearly, $\mu$ remains finite in the limit $b_1\to\pm\infty$, so
that $\exp(-\mu/2)$ does not damp the marginal distribution of $b_1$.
In fact, that damping comes from the factor of $1/\det Q=1/(1+b_1^2)$.
As a result, $b_1$ has well defined mean but its variance is undetermined.
Thus, estimators of $b_1$ evade the central limit theorem.

Nevertheless, the situation is pretty well under control.  The appropriate
question to ask of a distribution such as that in eq.\ (\ref{distb})
in the context of parameter estimation is not the value of the variance,
but an appropriate measure of the variation in the estimate. One could
quote the width at half maximum, or the limits such that 68\% of the
probability lies within these limits. Along with this one asks, if we
make $N$ measurements of $b_1$ then how does such a measure of variation
change with $N$.

A numerical investigation shows that when $\sigma_0=\sigma_1=1$, the modal
value is $b=0.345897$. Since the distribution is skew, the modal value and
the mean are different. The full width at half maximum is contained
in $-0.3485\le b_1\le 1.26641$, and this range contains 56.1\% of the
integral. The 68\% probability interval is $-0.575\le b_1\le1.38$.
Either of these ranges can be quoted as an estimate of the error in the
modal value.

To answer the question about the distribution of means, we use the
characteristic function.
If $f(x)$ is the distribution of $x$, then the Fourier transform
$\tilde f(x)$ is called the characteristic function. Since $f(x)$
is non-negative and integrable, being a probability distribution,
it is also square integrable, so that the characteristic function
exists. The characteristic function of the mean of $N$ numbers,
$\mu_N$, is
\beq
   \chi_N(\omega) = \int d\mu_N \exp(i\omega \mu_N)
       \int \left[\prod_{j=1}^N dx_i f(x_i)\right]
      \delta\left(\sum_{j=1}^N dx_i-N\mu_N\right)
    = \tilde f^N\left(\frac\omega N\right).
\label{charfn}\eeq
Fourier transforming this gives the distribution of $\mu_N$. While
this general method remains valid for the distributions ${\cal
P}^L_1$, above, it does not seem possible to perform the Fourier
transformations in closed form. So, instead of writing down an
impenetrable formula for the distribution of means, we investigate
useful subsidiary questions.

Define the skew of a distribution by
\beq
   {\cal S} = \frac{x_m}{\langle x\rangle} - 1
\label{skew}\eeq
where $x_m$ denotes the modal (most probable) value, and $\langle
x\rangle$ is the mean. The skew is nonzero for every skewed distribution,
being positive if the distribution is skewed to the left and negative
otherwise. For the distribution ${\cal P}^0_1$, we found ${\cal
S}=1.3\cdots$. For the distribution of means of $N$ values, $\cal S$
decreases. A Monte Carlo estimate for $\sigma_0 = \sigma_1 = 1$ indicates
that ${\cal S}\approx3.5/\sqrt{N}$. This estimate was obtained using
values of $N$ between 1 and 100.  A similar result was obtained for a
measure of skewness that compares the median and the mean in a manner
analogous to eq.\ (\ref{skew}).

At the median of a distribution, the cumulative distribution becomes
equal to 0.5. The errors on the median can be defined as the points
at which the cumulative distribution is either 0.34 above or below.
The distribution of the means of $N$ numbers narrows rapidly, and
we find in a Monte Carlo estimate that both these intervals decrease
as $1/\sqrt{N}$.

In conclusion, for the estimation of $b_1$ and confidence limits on the
estimate, it matters little that the central limit theorem does not hold.
The mean is well defined, and its difference with the mode and median
scale with a factor of $1/\sqrt{N}$. The 68\% confidence limits also
scale as $1/\sqrt{N}$. We therefore quote the mean and 68\% confidence
limits on it as estimators for the Pad\'e coefficients. These estimators
are easy to incorporate into jack-knife and bootstrap analyses.  In parts
of our analysis the estimators of the series coefficients are also not
Gaussian distributed; even so, the non-parametric statistical analysis
outlined here suffices.


\begin{thebibliography}{99}
\bibitem{cep}
   A.\ Barducci \etal, \pl, B 231 (1989) 463;\\
   M.\ A.\ Halasz \etal, \pr, D 58 (1998) 096007;\\
   J.\ Berges and K.\ Rajagopal, \np, B 538 (1999) 215.
\bibitem{pressure}
   R.\ V.\ Gavai and S.\ Gupta, \pr, D 68 (2003) 034506.
\bibitem{nt4}
   R.\ V.\ Gavai and S.\ Gupta, \pr, D 71 (2005) 114014.
\bibitem{biswa}
   C.\ R.\ Allton \etal, \pr, D 71 (2005) 054508.
\bibitem{others}
   Z.\ Fodor and S.\ Katz, \jhep, 0203 (2002) 014;\\
   C.\ R.\ Allton \etal, \pr, D 66 (2002) 074507;\\
   M.-P.\ Lombardo and M.\ d'Elia, \pr, D 67 (2003) 014505;\\
   Ph.\ de Forcrand and O.\ Philipsen, \np, B 642 (2002) 290;\\
   C.\ R.\ Allton, \etal, \pr, D 68 (2003) 014507;\\
   Z.\ Fodor and S.\ Katz, \jhep, 0404 (2004) 050;\\
   Ph.\ de Forcrand and O.\ Philipsen, \jhep 0701 (2007) 077;\\
   C.\ Bernard \etal, \pr, D 77 (2008) 014503.
\bibitem{misha06}
   M.\ A.\ Stephanov, \pr, D 73 (2006) 094508.
\bibitem{gottlieb}
   S.\ A.\ Gottlieb \etal, \prl, 59 (1987) 2247.
\bibitem{baker}
   G.\ A.\ Baker and P.\ Graves-Morris, {\sl Encyclopedia of Mathematics:
   Pad\'e Approximants\/}, Vol 13, Part 1, Addison-Wesley Publishing
   Company, Reading, Massachusetts, (1981).
\bibitem{hotqcd}
   Y.\ Aoki \etal, \pl, B 643 (2006) 46;\\
   M.\ Cheng \etal, \pr, D 74 (2006) 054507;\\
   C.\ Detar \etal, {\sl PoS\/} LAT2007 (2007) 179.
\bibitem{sewm}
   R.\ V.\ Gavai and S.\ Gupta, \np, A 785 (2007) 18.
\bibitem{quenched}
   R.\ V.\ Gavai and S.\ Gupta, \pr, D 67 (2003) 034501.
\bibitem{structure}
   R.\ V.\ Gavai and S.\ Gupta, \pr, D 72 (2005) 054006.
\bibitem{fluct}
   M.\ Asakawa, U.\ Heinz and B.\ Muller, \prl, 85 (2000) 2072;\\
   S.\ Jeon and V.\ Koch, \prl, 85 (2000) 2076.
\bibitem{critexpt}
   M.\ A.\ Stephanov, K.\ Rajagopal and E.\ V.\ Shuryak, \pr, D 60 (1999) 114028;
\\
   M.\ A.\ Stephanov, K.\ Rajagopal and E.\ Shuryak, \prl, 81 (1998) 4816;\\
   G.\ S.\ F.\ Stephans, nucl-ex/0607030;\\
   P.\ Sorensen (STAR Collaboration), nucl-ex/0701028.
\bibitem{mariapade}
   M.\ P.\ Lombardo, hep-lat/0509181.
\bibitem{koch}
   V. Koch, A.\ Majumder and J.\ Randrup, \prl, 95 (2005) 182301.
\bibitem{linkage}
   R.\ V.\ Gavai and S.\ Gupta, \pr, D 73 (2006) 014004.
\bibitem{cooperfrye}
   F.\ Cooper and G.\ Frye, \pr, D 10 (1974) 186.
\bibitem{largen}
   L.\ McLerran and R.\ D.\ Pisarski, \np, A 796 (2007) 83.
\bibitem{pade}
   H.\ Pade, {\sl Ann.\ de l'Ecole Norm.\ Sup.\ $3^{eme}$ Serie\/},
   9, Suppl. (1892) 3.
\bibitem{gaunt}
  D.\ S.\ Gaunt and A.\ J.\ Guttmann, p.\ 181, {\sl Phase Transitions
  and Critical Phenomena\/}, Vol. 3, eds.\ C.\ Domb and M.\ S.\ Green,
  Academic Press, London, 1974.
\bibitem{gaussrat}
   R.\ C.\ Geary, {\sl J.\ Royal Stat.\ Soc.\/}, 93 (1930) 442.
\end{thebibliography}
\end{document}